\documentclass[twocolumn]{aastex701}

\usepackage{hyperref}
\usepackage{amsmath}
\usepackage{amssymb}
\usepackage{xspace}

\newcommand{\DSalpha}{\texttt{DSalpha}\xspace}
\newcommand{\DSgamma}{\texttt{DSgamma}\xspace}
\newcommand{\DSepsilon}{\texttt{DSepsilon}\xspace}
\newcommand{\DSnu}{\texttt{DSnu}\xspace}
\newcommand{\DSTBB}{\texttt{T\_BB}\xspace}
\newcommand{\prospector}{\texttt{Prospector}\xspace}

\begin{document}

\title{  The \^G Search for Extraterrestrial Civilizations with Large Energy Supplies. V. When Galaxies Glow with Industry}



\author[orcid=0000-0002-0212-4563,sname='North America']{Olivia Curtis}
\affiliation{Department of Astronomy and Astrophysics, The Pennsylvania State University, 251 Pollock Road, University
Park, PA 16802, USA}
\affiliation{Penn State Extraterrestrial Intelligence Center, 525 Davey Laboratory, 251 Pollock Road, Penn State, University Park, PA, 16802, USA}
\affiliation{Institute for Gravitation and the Cosmos, The Pennsylvania State University, University Park, PA 16802, USA}
\email[show]{ocurtis@psu.edu}

\author[orcid=0009-0006-3368-6469,sname='North America']{Aidan J. Rowland}
\affiliation{Department of Astronomy and Astrophysics, The Pennsylvania State University, 251 Pollock Road, University
Park, PA 16802, USA}
\affiliation{Penn State Extraterrestrial Intelligence Center, 525 Davey Laboratory, 251 Pollock Road, Penn State, University Park, PA, 16802, USA}
\email[]{ajr7476@psu.edu}

\author[orcid=0000-0001-6160-5888,sname='North America']{Jason T. Wright}
\affiliation{Department of Astronomy and Astrophysics, The Pennsylvania State University, 251 Pollock Road, University
Park, PA 16802, USA}
\affiliation{Penn State Extraterrestrial Intelligence Center, 525 Davey Laboratory, 251 Pollock Road, Penn State, University Park, PA, 16802, USA}
\affiliation{Center for Exoplanets and Habitable Worlds, The Pennsylvania State University, 525 Davey Laboratory, 251 Pollock Road, Penn State, University Park, PA 16802, USA}
\email[]{astrowright@gmail.com}

\author[0000-0001-6842-2371]{Caryl Gronwall}
\affiliation{Department of Astronomy and Astrophysics, The Pennsylvania State University, 251 Pollock Road, University
Park, PA 16802, USA}
\affiliation{Institute for Gravitation and the Cosmos, The Pennsylvania State University, University Park, PA 16802, USA}
\email[]{cag18@psu.edu}

\author[0000-0003-4337-6211]{Jakob M. Helton}
\affiliation{Department of Astronomy and Astrophysics, The Pennsylvania State University, 251 Pollock Road, University
Park, PA 16802, USA}
\email{jakobhelton@psu.edu}

\author[0000-0001-6755-1315]{Joel Leja}
\affiliation{Department of Astronomy and Astrophysics, The Pennsylvania State University, 251 Pollock Road, University
Park, PA 16802, USA}
\affiliation{Institute for Gravitation and the Cosmos, The Pennsylvania State University, University Park, PA 16802, USA}
\affiliation{Institute for Computational and Data Sciences, The Pennsylvania State University, University Park, PA 16802, USA}
\email{joel.leja@psu.edu}

\begin{abstract}

We present the most robust stellar population synthesis (SPS)-based search for galaxy-spanning technological waste heat to date, applied to 129 nearby galaxies spanning a wide range of spectral energy distribution (SED) types, including ultraluminous IR galaxies and MIR-luminous active galactic nuclei (AGN). We incorporate the AGENT Dyson sphere formalism into the Flexible Stellar Population Synthesis code at the stellar population level, so nebular and dust emission respond self-consistently to Dyson sphere reprocessing. With \texttt{Prospector}, we perform a suite of 1,419 injection recovery tests across a range of covering fractions, $\alpha$, where we successfully recover the injected covering fractions (best-fit slope $m = 0.92$) and detect them through Bayesian model selection down to $\alpha \sim 4$--$5\%$ in quiescent galaxies. None of our 129 galaxies prefer a Dyson sphere component, and we place the first per-galaxy 95\% upper limits on warm ($T_{\rm BB} \gtrsim 100$K) swarms, reaching a median $\alpha < 0.3\%$ across quiescent hosts without a dominant AGN. Our injection-calibrated detection rates convert these zero detections into a population bound of $<2.6\%$ of galaxies hosting $\alpha = 25\%$ swarms ($95\%$ confidence). Because survey colors cannot separate waste heat from starbursts and AGN, we develop a scaffold for future searches, running from inexpensive archival screens such as the Balmer decrement and the stellar-to-dynamical-mass offset a swarm leaves behind, through resolved fitting with nuclear excision, to PRIMA FIR photometry that makes targeted JWST imaging decisive. We find that the outskirts of quiescent galaxies are the best hunting grounds for future technosignature searches.
  
\end{abstract}

\keywords{\uat{Search for extraterrestrial intelligence}{2127} --- \uat{Technosignatures}{2128} --- \uat{Galaxies}{573} --- \uat{IR excess galaxies}{789}}

\section{Introduction}

\cite{Dyson1960} and \cite{Slysh1985} showed that any sufficiently advanced civilization that has outsourced its technology to space will, according to the laws of thermodynamics \citep{Wright2023}, produce waste heat in the form of IR and microwave emissions. Many authors (e.g., \citealt{Wright2014a}; \citealt{Cirkovic2018}) since then have hypothesized that a space-faring species would settle much or most of its host galaxy, eventually collecting and using enough starlight that they would affect the energy balance of the galaxy on their way to being a Type~{\sc iii} civilization on the Kardashev scale \citep{Kardashev1964}. The collection of starlight via swarms of orbiting solar panels is called a ``Dyson sphere,'' but note that this does not imply collection would happen via a single solid object, spherical or otherwise. Any collection of circumstellar technology (e.g., a satellite swarm) is a ``Dyson sphere'' \citep{Wright2020}.

In their pioneering work, \cite{Annis1999} set the first upper limit for technological waste heat that is extragalactic in origin by looking for galaxies whose radial surface brightness profiles deviate significantly from well-established scaling relations \citep{Tully1977, Zackrisson2015}. \cite{Griffith2015} pushed these upper limits further by searching N$\sim10^5$ extended sources in the Wide-field IR Survey Explorer (WISE; \citealt{Wright2010}) for mid-infrared (MIR) excesses consistent with technological waste heat \citep{Wright2014b, Wright2014a}, showing that waste heat from technosignatures and dust have very different properties in galaxy spectral energy distributions (SEDs), spectra, and imagery. While \cite{Griffith2015} succeeded in identifying anomalously MIR-bright galaxies, namely a population of passive red spiral galaxies \citep{Valentijn1990}, their search strategy was limited to resolved sources in the WISE catalog and relied on color cuts as their distinguishing factor. Their upper limits were conservative, corresponding to covering fractions of $\alpha \gtrsim 50$--$80\%$, set by the presence of ordinary dusty and starburst galaxies, and AGN.

Within the Milky Way, targeted searches for stellar-scale Dyson spheres have established that such structures are rare \citep{Suazo2022, Suazo2024, Korn2026, Zackrisson2026}. Colonization-timescale arguments \citep[cf.][]{Hart1975, Tipler1980, Gray2015} nonetheless suggest that a spacefaring species could restructure the radiative energy balance of an entire galaxy \citep{Cirkovic2018}, and \cite{Kardashev1964} noted that the energy use of such a Type~{\sc iii} civilization should be detectable across cosmological distances, making the extragalactic regime a natural complement to stellar-scale searches.

\cite{lacki2016} pushed this picture to its extreme with the ``blackbox,'' an entire galaxy enclosed behind an opaque screen that radiates just above the temperature of the cosmic microwave background, where a heat engine reaches its maximum thermodynamic efficiency. No source in the Planck Catalog of Compact Sources carries that spectrum. The null result covers roughly three million galaxies and reaches Milky Way analogs out to a comoving $700$~Mpc, and it led Lacki to conclude that Type~{\sc iii} societies effectively do not exist in the observable Universe. Those limits, however, only constrain galaxies that are completely cloaked at very low temperatures. 

Subsequent theoretical work by \cite{lacki2019} then showed that this all-or-nothing picture is too restrictive as most of a galaxy's luminosity comes from its brightest stars, so a civilization need not enclose every star to leave a detectable photometric signature. Low-mass stars are, moreover, far more numerous and far longer-lived than luminous stars, which makes them the more natural targets for a civilization seeking a stable, long-lived energy supply. Selectively cloaking lower-luminosity stars below some luminosity threshold produces galaxies with anomalously red near- and MIR colors, as well as unexpectedly blue ultraviolet emission in star-forming systems. However, \cite{lacki2019} relies on the same crude color cuts that limited the original \^G survey \citep{Wright2014b, Wright2014a}.


More recently, \cite{Huang2026} used data from the Wide-field IR Space Explorer (WISE; \citealt{Wright2010}) and CatWISE \citep{Eisenhardt2020} photometry of $\sim2\times10^4$ nearby galaxies to place upper limits on waste heat luminosities, quoting covering fractions below $1.7$--$2.9\%$ across blackbody temperatures of $150$--$600$K. These caps are not measurements of any individual galaxy's covering fraction, however. The method treats the entire observed W3/W4 flux (plus $3\sigma$) as potentially technological, subtracting no stellar, dust, or AGN emission, and the quoted $\alpha$ values divide the sample-median luminosity cap by a fiducial Milky-Way stellar luminosity rather than each galaxy's own. Every galaxy with AGN- or starburst-like MIR (MIR) colors is furthermore masked out of the sample, removing exactly the systems in which waste heat is hardest to distinguish from astrophysical emission (a cut that, as the authors note, could also remove a genuine Type~{\sc iii} host). Such an approach can only cap fluxes, so it cannot detect a waste heat component, identify candidates, or address the degeneracy between technological waste heat and interstellar dust.

Still, these studies laid the groundwork for the approach taken here. Through detailed modeling of galaxies' brightnesses at a range of wavelengths, one can determine a galaxy's inventory of stars, gas, dust, active black holes, and Dyson spheres. SED modeling improves the search in three specific ways. First, Dyson spheres affect all parts of a galaxy's SED, simultaneously building a mid-IR bump while dimming the UV--optical starlight that powers it, and only a fit to the full SED can model this exchange in an energy-conserving fashion. Second, the UV-through-near-IR photometry independently fixes the stellar mass, star formation rate, and attenuation, which predicts the stellar-heated dust emission and exposes any genuine excess, rather than permitting all MIR flux to be attributed to dust, or indeed to waste heat. Third, Bayesian model comparison converts the search from population-level color cuts into a per-galaxy measurement, yielding calibrated posteriors on the covering fraction and temperature of a putative swarm and a model-selection statistic that identifies candidates. The same physical reasoning extends beyond photometry; a MIR excess produced by ordinary dust must be accompanied by correspondingly high nebular reddening (i.e., the Balmer decrement), whereas waste heat requires none, giving the SED-based search a spectroscopic follow-up path that no color-based method possesses.

This paper is the fifth in the \^G series \citep{Wright2014b, Wright2014a, Griffith2015, Wright2016}. Those papers modeled host galaxies with fixed templates and selected candidates by color, with their conclusions calling for more sophisticated methodologies that are capable of reaching smaller waste heat luminosities. We provide those very methodologies by building on these studies in the style of \cite{lacki2019}, where, instead of using hot galaxy templates, we instead fit the stellar, dust, and AGN emission of every galaxy instead. We thus present the most robust stellar population synthesis (SPS)-based search for extragalactic waste heat that has been conducted to date, showing that we can be sensitive to Type~{\sc iii} species that are using as little as $\sim4$--$5\%$ of the starlight of their host galaxies. This is an order of magnitude better than the color cut limits of \cite{Griffith2015} and, unlike the flux caps of \cite{Huang2026}, is inferred for each galaxy individually with its stellar, dust, and AGN emission modeled jointly. Ancillary to these results, we also end up ranking every galaxy in the \cite{Brown2014} galaxy atlas by how anomalously IR they are, and we briefly discuss the impacts that such a catalog, when extended to a much larger sample size, could have on the entire field of extragalactic astronomy.

\cite{Wright2014b} supplies the exact formalism we need (i.e., the AGENT formalism; see~\ref{sec:AGENT}), which we incorporate into state-of-the-art SPS models \citep[see, e.g.,][for a review]{Conroy2013} to characterize the SEDs of 129 nearby galaxies \citep{Brown2014} from the ground up, incorporating the AGENT parameters \citep{Wright2014a} directly into the Flexible Stellar Population Synthesis (\texttt{FSPS}) code at the stellar population level. This yields a self-consistent forward model in which Dyson sphere photons are injected before light propagates into the surrounding gas and dust, allowing posterior probability distributions to be derived for the covering fraction and temperature of a swarm. We pair this forward model with the \prospector Bayesian inference framework \citep{leja2017, Johnson2021} to fit 24-band photometry from the \citet{Brown2014} galaxy atlas. 

We search for galaxies with MIR fluxes too high to be explained by the star formation, asymptotic giant branch (AGB) stars, dust, and AGN that the rest of their SEDs require them to have. Previous SED fitting studies (e.g., \citealt{leja2017,leja2018}) find MIR residuals below 10\% once an AGN component is included. Galaxies hosting Dyson spheres produce WISE colors similar to MIR-bright AGN \citep{Wright2014b, Griffith2015}, but our Bayesian framework disentangles these using the full multiwavelength SED rather than color cuts alone, and spatial information provides a further discriminant (\S~\ref{sec:resolvedMethod}). This approach also allows us to perform resolved searches for Dyson sphere emission, which lets us break centralized degeneracies and search for Type~{\sc iii} civilizations that have spread to only part of their host galaxy \citep{Wright2021}.

We proceed with a discussion of the AGENT formalism for Dyson sphere reprocessing in \S~\ref{sec:AGENT}, before discussing our observations in \S~\ref{sec:galCat}. We then describe the basics of SPS modeling in \S~\ref{sec:SPS}, the \prospector Bayesian inference framework \citep{leja2017, Johnson2021} in \S~\ref{sec:bayes}, and the framework behind resolved searches in \S~\ref{sec:resolvedMethod}. We present results in \S~\ref{sec:results}, discussing our injection recovery tests (\S~\ref{sec:injectRecov}), the Bayesian Information Criterion we use for model selection (\S~\ref{sec:modelSelection}), how Dyson sphere emission affects AGN parameters (\S~\ref{sec:AGNParams}) and star formation histories (SFHs) (\S~\ref{sec:SFH}), the effects of using a limited, survey-telescope only filter set (\S~\ref{sec:limitedFilters}), and using resolved SED fitting to break the Dyson sphere and hot AGN degeneracy (\S~\ref{sec:resolved}).

We discuss our results in \S~\ref{sec:discussion}, covering our principal results (\S~\ref{sec:principal}), the two signatures a waste-heat-free model leaves when a swarm goes unmodeled (\S~\ref{sec:failuremodes}), and the resolved fitting that breaks the AGN degeneracy (\S~\ref{sec:agndegen}). We then comment on our future plans to search the $\sim10^7$ galaxies with archival MIR photometry and to nearby resolved galaxies (\S~\ref{sec:scaling}), and we show how the proposed PRobe FIR Mission for Astrophysics (PRIMA; \citealt{Glenn2025}) makes targeted JWST/MIRI imaging decisive (\S~\ref{sec:futureobs}). We close with a tiered strategy for confirming a candidate (\S~\ref{sec:confirming}), running from the Balmer decrement and the infrared-excess versus ultraviolet-optical color plane to the MIR spectroscopy that settles the case. We then rank the \cite{Brown2014} galaxy atlas by how anomalously MIR bright they are in \S~\ref{sec:anomalousTable} before concluding in \S~\ref{sec:conclusions}. Throughout, magnitudes are quoted in the AB system, and we adopt the \cite{Planck2018} cosmology.

\section{Methodology}

\subsection{The AGENT Formalism}
\label{sec:AGENT}

\cite{Wright2014a} parametrizes the absorption and re-emission of stellar light by Dyson spheres following the AGENT (or $\alpha\gamma\epsilon\nu' T_{\rm BB}$) formalism. Here, some fraction $\alpha$ of the total stellar luminosity of the galaxy is absorbed by obscuring bodies, and some fraction $\gamma$ of the total stellar luminosity is re-emitted as a blackbody of temperature $T_{\rm BB}$. The other two parameters account for extraneous energy that is either being generated ($\epsilon$) or re-emitted ($\nu'$) by the swarm elements. The formalism implicitly assumes that swarms are distributed isotropically, with no preferential placement into or out of our line of sight, which is reasonable for an ensemble of many independently oriented circumstellar swarms. Conservation of energy and a steady state assumption thus requires

\begin{equation}
    \alpha + \epsilon = \gamma + \nu'.
\end{equation}

\noindent In general, we can assume negligible non-thermal losses or emissions or non-stellar sources of luminosity ($\epsilon = \nu'=0$) such that the amount of stellar light absorbed is equal to the amount emitted as a blackbody (i.e., $\alpha=\gamma$). 




\subsection{Galaxy catalog}
\label{sec:galCat}

We use as a test sample the SEDs in the \cite{Brown2014} galaxy atlas. This catalog provides matched-aperture photometry for 129 galaxies across 24 photometric bands from the Galaxy Evolution Explorer (GALEX; \citealt{Morrissey2007}), the \textit{Swift} UV/optical monitor telescope (UVOT; \citealt{Roming2005}), the Sloan Digital Sky Survey III (SDSS; \citealt{Aihara2011}), the Two Micron All Sky Survey (2MASS; \citealt{Skrutskie2006}), WISE \citep{Wright2010}, and the \textit{Spitzer} space telescope \citep{Werner2004}. For the bright galaxies ($m_{\rm FUV}<18$ and $m_{K_s}<13$), \cite{Brown2014} assumes an uncertainty floor of 0.05~mag for the SDSS and 2MASS bands and a floor of 0.10~mag for the other bands. For the fainter galaxies in the sample, \cite{Brown2014} estimated uncertainties and background errors from the fluxes at 24 locations surrounding the galaxy. Since most of these galaxies are bright, nearby galaxies, most of the flux measurements have uncertainties of order $\sim0.1$ mag. The atlas spans the full range of morphological types present in the local universe (i.e., quiescent ellipticals, star-forming spirals, irregular and interacting galaxies, and galaxies hosting AGN), which makes it an ideal test bed for characterizing how our model behaves across galaxy morphology and dust content.

We deliberately restrict our fits to the matched-aperture photometry of \cite{Brown2014} so that every galaxy in the sample has uniform and self-consistent multiwavelength coverage. Archival FIR photometry from Herschel, together with data from ALMA and JWST, exists for many of these galaxies, but we do not incorporate it here for two reasons. First, it is not part of the \cite{Brown2014} matched-aperture catalog, and folding it in would require careful aperture matching across facilities with very different angular resolutions and point spread functions, a step we deliberately avoid so that aperture-matching systematics do not become a large and uncontrolled part of this study. Second, we want our covering-fraction limits to be set by a single homogeneous photometric system rather than by heterogeneous archival coverage that varies from galaxy to galaxy. We note, however, that this archival coverage would be valuable for follow-up. Herschel broadband photometry in particular, despite its limited sensitivity, samples the cold-dust FIR continuum that most directly constrains the temperature and amplitude of any waste-heat blackbody, so incorporating it into targeted follow-up of individual candidates would meaningfully tighten the limits derived here.

\begin{table}[!h]
    \footnotesize
    \centering
    \caption{The filters used in this work (in order of increasing wavelength), as well as their corresponding effective wavelengths and FWHM bandwidths.}
    \label{tab:filterlist}
    \begin{tabular}{c c c}
        \hline
        Filter & $\lambda_{\rm eff}$ ($\mu$m) & FWHM bandwidth ($\mu$m) \\ \hline
        GALEX \textit{FUV}            & 0.1531 & 0.0227 \\
        UVOT $W2$              & 0.2026 & 0.0557 \\
        UVOT $M2$              & 0.2238 & 0.0511 \\
        GALEX $NUV$            & 0.2286 & 0.0795 \\
        UVOT $W1$              & 0.2598 & 0.0681 \\
        UVOT $U$               & 0.3459 & 0.0557 \\
        SDSS $u$               & 0.3551 & 0.0582 \\
        SDSS $g$               & 0.4681 & 0.1262 \\
        UVOT $V$               & 0.5419 & 0.0730 \\
        SDSS $r$               & 0.6165 & 0.1149 \\
        SDSS $i$               & 0.7480 & 0.1238 \\
        SDSS $z$               & 0.8931 & 0.0994 \\
        2MASS $J$              & 1.232  & 0.215 \\
        2MASS $H$              & 1.644  & 0.263 \\
        2MASS $K_s$            & 2.159  & 0.279 \\
        WISE $W1$              & 3.357  & 0.793 \\
        \textit{Spitzer} [3.6] & 3.544  & 0.743 \\
        \textit{Spitzer} [4.5] & 4.487  & 1.010 \\
        WISE $W2$              & 4.606  & 1.106 \\
        \textit{Spitzer} [5.8] & 5.710  & 1.391 \\
        \textit{Spitzer} [8.0] & 7.841  & 2.831 \\
        WISE $W3$              & 11.81  & 8.67 \\
        WISE $W4$              & 22.14  & 4.40 \\
        \textit{Spitzer} [24]  & 23.51  & 5.03 \\ \hline
    \end{tabular}
\end{table}

Table~\ref{tab:filterlist} delineates the filters, effective wavelengths, and FWHM bandwidths used throughout this work. We also run a parallel set of fits using only GALEX FUV/NUV, SDSS $ugriz$, 2MASS $JHK_s$, and WISE W1--W4, omitting Swift UVOT and \textit{Spitzer} data in order to assess sensitivity under the conditions of an all-sky search where only wide-field survey data are available (see \S~\ref{sec:limitedFilters}).

For the injection tests, we use Equation~\ref{eq:AGENT2} to generate 11 copies of each SED with $\alpha = \gamma \in \{0, 1, 2, 3, 4, 5, 10, 15, 20, 25, 50\}\%$ and $T_{\rm BB} = 300$K, for a total of 1,419 injected SEDs. The choice of $T_{\rm BB} = 300$K is both physically motivated and observationally convenient, and we discuss this choice, along with how the swarm temperature sets which astrophysical component is most easily confused with the waste heat, in Appendix~\ref{sec:Tappendix}. Because we leave $T_{\rm BB}$ free in every fit, this choice fixes only the temperature of the injected signals and not the temperature recovered for any galaxy.

We note that na\"{i}vely applying Equation~\ref{eq:AGENT2} to the full SED rather than only the stellar component would not be physical, since dust emission, nebular emission, and black hole accretion disk emission are also modulated even though Dyson spheres do not directly intercept them. As such, in the next section, we take the robust approach of only applying the AGENT formalism to individual stellar populations.

\subsection{Stellar Population Synthesis}
\label{sec:SPS}


The first step in modeling a galaxy's SED involves the construction of a simple stellar population (SSP), which is a collection of all stars of varying ages and metallicities that inhabit a galaxy at some time \citep{Tinsley1972, Searle1973, Larson1978, Conroy2009, Walcher2011}. The primary method of doing this \citep{Charlot1991} involves combining stellar isochrones, an initial mass function (IMF; $\phi(M)$), and stellar spectral templates to produce a spectral luminosity (in units $\rm{erg} \; s^{-1} \; M_\odot^{-1}$) as

\begin{equation}
    \label{eq:SSP}
    L_\nu^{\rm SSP}(t,Z) = \int_M\phi(M)_{t,z}L_\nu(M,t,Z) ,
\end{equation}

\begin{figure*}[!tp]
    \centering
    \includegraphics[width=0.62\textwidth]{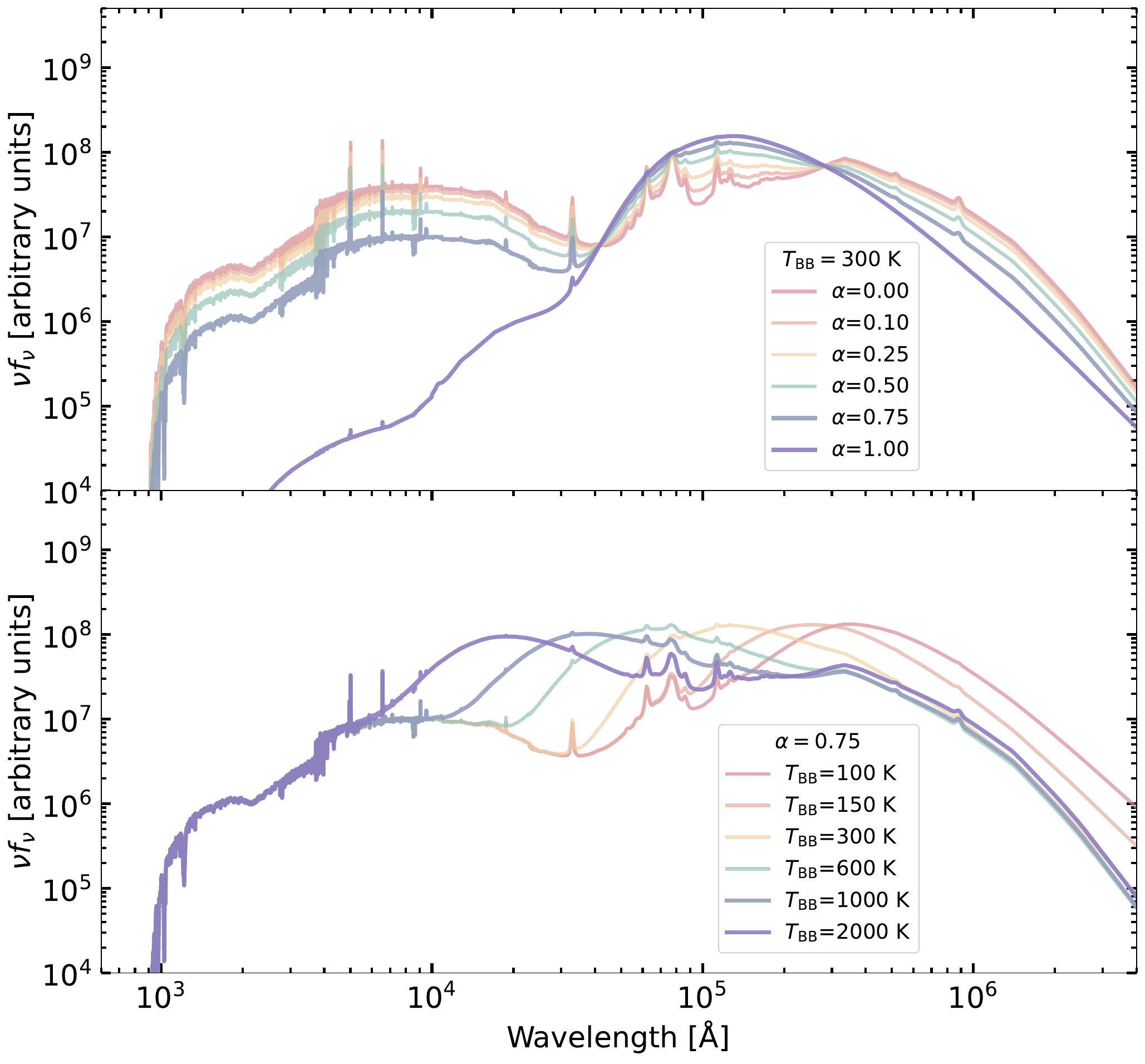}
    \caption{Spectral energy distributions ($\nu f_\nu$, arbitrary units) vs. wavelength for a typical star-forming galaxy produced with \texttt{python-fsps}. In the top panel, we fix \DSTBB at $300$K and increase the Dyson sphere covering fraction \DSalpha from $0$ to $1$. At $\alpha = 0$ (pink line), the galaxy exhibits the usual rising UV--optical stellar continuum followed by molecular dust absorption and thermal emission in the mid-to-far IR, including prominent PAH features near $10^5$~\AA. As $\alpha$ increases, more UV--optical light is absorbed and re-emitted as a blackbody, peaking near $\sim10^5$~\AA. Note that, as Dyson sphere coverage grows, the PAH features around $10^5$~\AA\ also diminish since fewer UV photons are available to excite the surrounding dust. Even when $\alpha = 1$ and every star is fully covered, some faint dust emission persists as the $300$K Dyson sphere photons continue to heat the surrounding dust at low levels. In the bottom panel we instead hold \DSalpha at $0.75$ and vary \DSTBB from $100$ to $2000$K, showing how the waste-heat bump marches across the IR range as the swarm temperature changes. A cold $100$K swarm peaks in the far IR beyond $\sim10^6$~\AA, near where cold interstellar dust emits, whereas a hot $2000$K swarm peaks in the near-to-mid IR near $\sim10^4$--$10^5$~\AA, overlapping both the stellar Rayleigh-Jeans tail and warm AGN dust. This self-consistent coupling between the swarm and the galaxy's gas and dust is the key advantage of injecting AGENT emission at the stellar population level.}
    \label{fig:FSPSTest}
\end{figure*}

\noindent where $L_\nu(M,t,Z)$ is the SED of a star given its mass ($M$), age ($t$), and metallicity ($Z$; \citealt{Walcher2011}). Throughout, we use a \cite{Kroupa2001} IMF, MIST isochrones \citep{Dotter2016, Choi2016}, and the C3K spectral library \citep{Conroy2009, Conroy2013}.

However, the total SED of a galaxy is more than just the sum of its stars, instead consisting of a complex interplay of its dust, gas, geometry, star formation history, and accreting supermassive black hole. A complete description of the phenomena that go into the construction of a complex stellar population can be found in \cite{Walcher2011}, \cite{Conroy2013}, or \cite{leja2017}, but we summarize the key free parameters in Table~\ref{tab:FSPS}. We make use of the \texttt{FSPS} \citep{Conroy2009}\footnote{\url{https://github.com/cconroy20/fsps}} code and its corresponding Python wrapper \texttt{python-fsps}\footnote{\url{https://github.com/dfm/python-fsps}}. We modify \texttt{FSPS} and \texttt{python-fsps} to accept AGENT parameters by applying the AGENT formalism to Equation~\ref{eq:SSP} immediately after the SSP is generated. That is, given $L_\nu^{\rm SSP}$ in units of $L_\odot \; \rm{Hz}^{-1}$, its starlight after being reprocessed by Dyson spheres is

  \begin{multline}
    \label{eq:AGENT}
    L_\nu^{\rm SSP,DS}(t,Z) = (1-\alpha)L_\nu^{\rm SSP}(t,Z) \\
    + \left[ \gamma\frac{B_\nu(T_{\rm BB})}{\int B_\nu(T_{\rm BB})d\nu} + \nu'\frac{L_\nu^{\rm nt}}{\int L_\nu^{\rm nt}d\nu} \right] \int L_\nu^{\rm SSP}(t,Z)d\nu \;,
  \end{multline}

\noindent where the un-primed $\nu$ is frequency, $B_\nu(T_{\rm BB})$ is the Planck function at $T=T_{\rm BB}$ in units of $\rm{erg} \; s^{-1} \; cm^{-2} \; Hz^{-1} \; sr^{-1}$, and $L_\nu^{\rm nt}$ is the spectrum of some non-thermal energy disposal mechanism that is assumed to be electromagnetic in units of $L_\odot \; \rm{Hz}^{-1}$. We divide each of the two emission terms by its own integral over frequency, which keeps the shape of each spectrum but scales it so that it integrates to one. For the Planck function that integral evaluates to $\int B_\nu(T_{\rm BB})d\nu = \sigma_{\rm SB}T_{\rm BB}^4/\pi$ for a specific intensity, where $\sigma_{\rm SB}$ is the Stefan-Boltzmann constant. The bracketed kernel therefore carries units of $\rm{Hz}^{-1}$ and integrates to exactly $\gamma+\nu'$, so the reprocessing conserves the bolometric luminosity of the population whenever $\gamma+\nu'=\alpha$, and exceeds it by $\epsilon$ when the civilization disposes of energy drawn from outside the stellar population (e.g., from accretion onto black hole; \citealt{hsiao2021}; \citealt{Curtis2026}). For the rest of this manuscript, we set $\epsilon=\nu'=0$ and we set $\gamma=\alpha$ such that Equation~\ref{eq:AGENT} becomes

  \begin{multline}
      \label{eq:AGENT2}
      L_\nu^{\rm SSP,DS}(t,Z) = (1-\alpha)L_\nu^{\rm SSP}(t,Z) \\
      + \left[ \frac{\alpha \pi B_\nu(T_{\rm BB})}{\sigma_{\rm SB}T_{\rm BB}^4} \right] \int L_\nu^{\rm SSP}(t,Z)d\nu \;.
  \end{multline}

\noindent This placement is physically motivated since Dyson spheres intercept stellar photons before they can excite nebular emission or heat the interstellar dust. As a result, the light that propagates outward to the surrounding gas and dust already carries the imprint of the swarm. This means that as $\alpha$ increases, the dust emission features are suppressed alongside the UV--optical continuum and the overall dust emission diminishes. The features in question are chiefly the polycyclic aromatic hydrocarbon (PAH) bands, which in this formalism are heated by photons of all wavelengths but predominantly by the UV light of young stars, owing to the shape of the attenuation law. Importantly, dust emission never vanishes entirely even at $\alpha = 1$, since the $300$K blackbody photons emitted by the Dyson spheres can themselves heat the surrounding dust at some level. Throughout the rest of this manuscript, the AGENT parameters as implemented in \texttt{Prospector} are sometimes referred to as \DSalpha, \DSgamma, \DSepsilon, \DSnu, and \DSTBB to avoid confusion with other parameters in \texttt{FSPS} that share similar labels. 

We note that Equation~\ref{eq:AGENT2} treats the swarm as cloaking all stars, sampling every star in proportion to its luminosity such that a fraction $\alpha$ of the total starlight is reprocessed. Our implementation also permits the reprocessing to be restricted to stars in a single evolutionary phase, in which case $L_\nu^{\rm gal}$ in the absorbed and re-emitted terms is replaced by the spectrum of the targeted stars and $\alpha$ becomes the covering fraction of the swarms around them. All of our fits assume the swarm reprocesses light from all stars, but we do employ the stellar-evolutionary-phase-restricted variant in the forward modeling that we perform while considering targeted cloaking strategies in \S~\ref{sec:confirming}.

\begin{figure*}[!tp]
    \centering
    \includegraphics[width=0.70\textwidth]{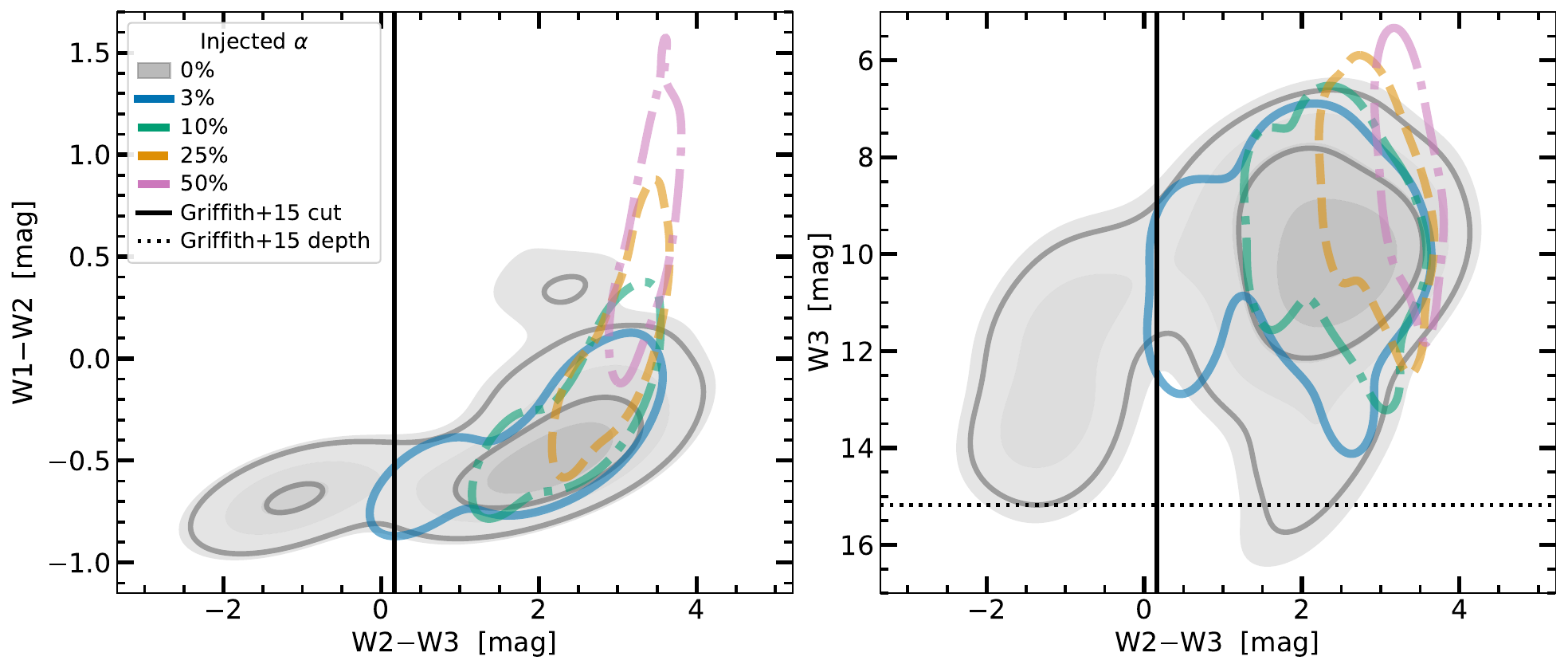}
    \caption{WISE color-color (\textit{left}: W1$-$W2 vs.\ W2$-$W3) and color-magnitude (\textit{right}: W3 vs.\ W2$-$W3) diagrams for all 129 galaxies in the \cite{Brown2014} atlas across five of our eleven Dyson sphere injection levels (see \S~\ref{sec:galCat}). In each panel, gray filled contours trace the baseline (0\%) population, while colored contours show the same 129 galaxies with Dyson emission injected at the labeled covering fraction. As $\alpha$ increases from 3\% to 50\%, the population contours shift progressively toward redder W2$-$W3 colors and brighter W3 magnitudes, reflecting the growing $\sim$300K blackbody contribution to the WISE W3 band. Even at $\alpha = 50\%$ the injected contours remain embedded in the color-space locus occupied by luminous IR galaxies (LIRGs) and warm-dust AGN. The solid black line marks the stellar-locus cut used by \cite{Griffith2015}, who removed all sources with W2$-$W3 $<$ 2 and W3$-$W4 $\leq$ 1 (Vega, converted here to AB) and required $\gamma$ to be $\geq 0.25$ for all of their candidates. The dotted black line in the right panel marks their effective detection limit ($\approx$ 10; Vega) in the WISE W3 band. 
    Previous extragalactic Dyson sphere searches that relied solely on WISE color selection \citep{Wright2014b, Griffith2015} could consequently place meaningful limits only at $\alpha \gtrsim 50$--80\% since DS-injected galaxies cannot be separated from these astrophysical populations by color cuts alone. Our full SED fitting approach, which simultaneously fits 24 photometric bands from the far-UV through the MIR, breaks this degeneracy and pushes the detection threshold an order of magnitude deeper.}
    \label{fig:WISEcolors}
\end{figure*}

\begin{deluxetable}{ll}
\tabletypesize{\scriptsize}
\tablecaption{Free parameters and priors for the \texttt{Prospector-$\alpha$} model used in this work. Nebular emission is included, with the gas ionization parameter held fixed and the gas-phase metallicity tied to the stellar metallicity, so neither is a free parameter.}
\tablenum{2}
\label{tab:FSPS}
\tablehead{\colhead{Parameter} & \colhead{Prior}}
\startdata
\texttt{logzsol}                      & Uniform: min=$-2.0$, max=$0.6$ \\
\texttt{logmass}                      & Uniform: min=7, max=12 \\
$r^1_{\rm SFH}\ldots r^6_{\rm SFH}$  & Student-$t$: mean=0, scale=0.3, DoF=2 \\
\texttt{duste\_umin}                  & Uniform: min=0.001, max=25 \\
\texttt{duste\_qpah}                  & Uniform: min=0.5, max=7 \\
\texttt{duste\_gamma}                 & Uniform: min=$10^{-5}$, max=1 \\
\texttt{dust\_index}                  & Uniform: min=$-2.0$, max=0.5 \\
\texttt{dust2}                        & Uniform: min=0, max=4 \\
\texttt{dust1}                        & Uniform: min=0, max=4 \\
\texttt{fagn}                         & LogUniform: min=$10^{-5}$, max=3 \\
\texttt{agn\_tau}                     & LogUniform: min=$10^{-3}$, max=150 \\
\DSalpha                              & Uniform: min=$10^{-3}$, max=1 \\
\DSTBB                                & Uniform: min=0.1K, max=1000K \\
\enddata
\end{deluxetable}

The top panel of Figure~\ref{fig:FSPSTest} illustrates this implementation, showing the SED of a typical star-forming galaxy as \DSalpha is varied from $0$ to $1$ with \DSTBB fixed at $300$K.  This figure demonstrates the dual-sided spectral imprint of Dyson spheres. That is, they suppress the UV--optical continuum while simultaneously building up a MIR blackbody excess, exactly the kind of spectral anomaly our fitting pipeline is designed to detect. The bottom panel shows the opposite---fixing \DSalpha to $0.75$ while sliding \DSTBB from $100$ to $2000$K to show how the shape of the SED changes as it reacts to different swarm temperatures.

Figure~\ref{fig:WISEcolors} illustrates why photometric color cuts alone cannot isolate Dyson sphere emission at the precision our framework achieves. Here, we show how WISE color-color and color-magnitude diagrams change after we perform the Dyson sphere injection that we describe in \S~\ref{sec:galCat}. The 0\% contours (gray) span a wide locus from quiescent ellipticals (blue, low W2$-$W3) to star-forming spirals and AGN hosts (red, high W2$-$W3 and W1$-$W2), reflecting the genuine diversity of the sample. Injecting DS emission shifts the entire distribution toward redder W2$-$W3 colors and brighter W3 magnitudes monotonically with $\alpha$, but the shift is subtle at low covering fractions since the 3\% contours are largely indistinguishable from the 0\% baseline by eye. At 25\% and 50\%, the separation becomes clear, yet the injected contours still overlap substantially with the astrophysical population, particularly with the color locus where IR luminous galaxies and warm-dust AGN naturally fall. \cite{Griffith2015} and \cite{Wright2014b} both presented WISE color-color and color-magnitude diagrams similar to these and, where they used the excess redness relative to the normal galaxy locus as their primary selection criterion, and their detection limits of $\sim$50--80\% reflect these overlapping contours. The UV-through-MIR SED fitting we perform below instead constrains the stellar mass, dust temperature distribution, star formation history, and AGN fraction independently of the MIR excess, which allows us to measure the Dyson sphere component directly rather than inferred from color alone.

\begin{figure*}[!tp]
    \centering
    \includegraphics[width=0.7\textwidth]{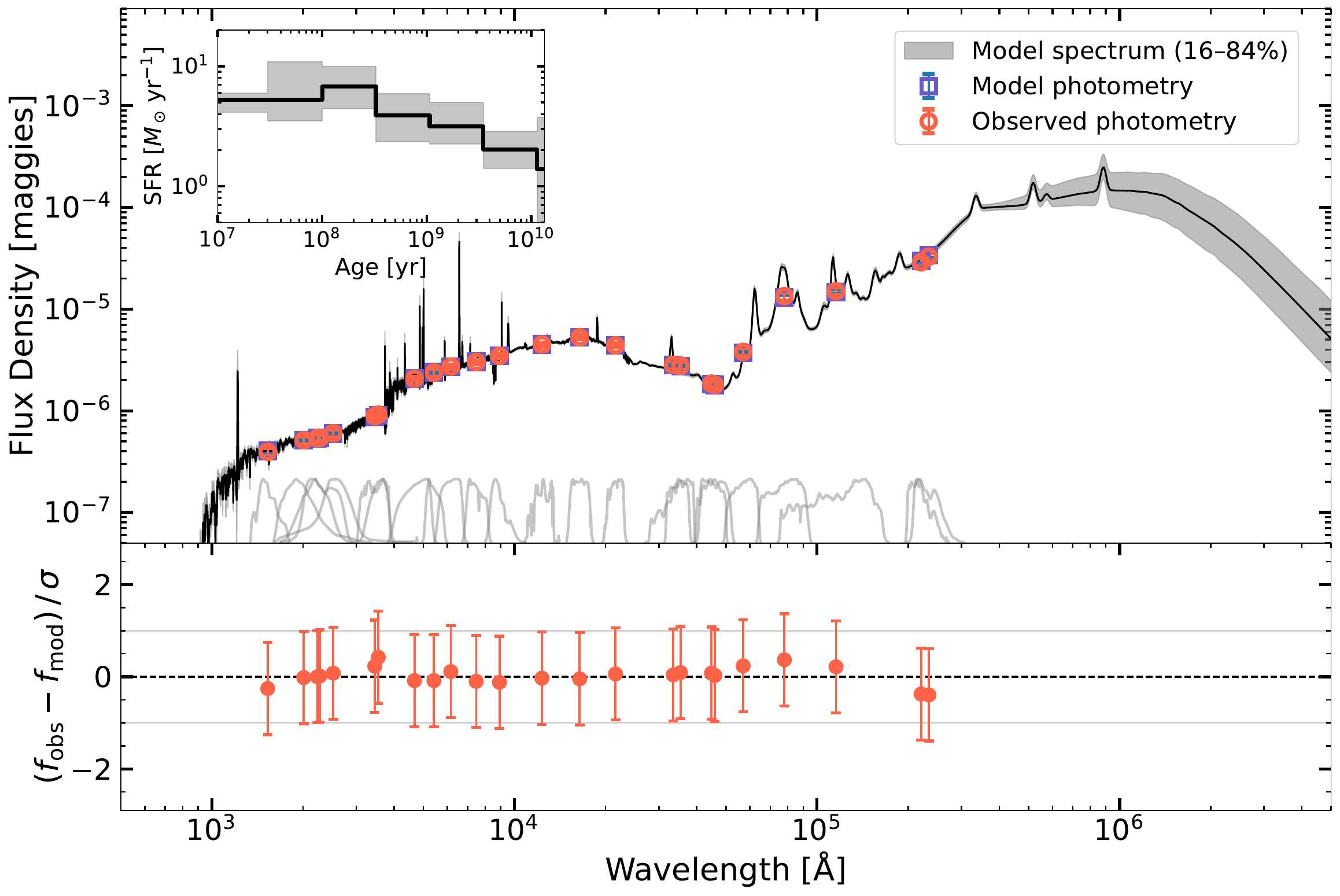}
    \caption{\textbf{Main panel:} Spectral flux density for the star-forming galaxy Arp~256~N fit with \prospector without AGENT parameters. The black curve and gray band show the median and 16--84\% credible interval of the model spectrum from 1,000 posterior draws while blue squares show model photometry and red circles show observed photometry. Gray curves at the bottom of the panel show the transmission profiles of all 24 photometric filters used in the fit. Wavelengths are plotted in the observed frame. The star formation history (SFH) inset shows the posterior median (black) and 16--84\% interval (gray band). The rising SFR toward recent times reflects the ongoing interaction and starburst activity in this system. As discussed in \S~\ref{sec:galCat}, \cite{Brown2014} inflated the reported photometric errors to 5--10\% to conservatively reflect survey systematics, which is why the residuals are all significantly $<1\sigma$. \textbf{Lower panel:} Normalized residuals $(f_{\rm obs} - f_{\rm mod})/\sigma$ for each photometric band and dashed line marks zero and light gray lines mark the $\pm 1\sigma$ level. The fit is representative of the atlas, reproducing the full ultraviolet-through-IR SED to well within the adopted uncertainties, and it is this baseline agreement against which an injected waste-heat excess must be detected.}
    \label{fig:ArpSpec}
\end{figure*}

\subsubsection{Bayesian Inference}
\label{sec:bayes}

The forward model described above is embedded in the \prospector \citep{leja2017, Johnson2021} inference framework, which wraps \texttt{FSPS} into a mature forward-modeling architecture. \prospector calls \texttt{FSPS} to generate a model SED, convolves it with the photometric filter transmission curves, and computes the likelihood by comparing to observed photometry. Throughout this work, we use a modified version of the \texttt{Prospector-$\alpha$} model of \cite{leja2017}, adopting the continuity star formation history prior of \cite{leja2019} but no mass--metallicity prior, for a total of 18 free parameters, including our 2 AGENT parameters, \DSalpha and \DSTBB.

Our priors are listed in Table~\ref{tab:FSPS}. Stellar metallicity (\texttt{logzsol}, in units of $\log_{10}(Z/Z_\odot)$) and total stellar mass (\texttt{logmass}) are both given broad uniform priors, with metallicity bounded by the model grid of the MIST isochrones \citep{Dotter2016, Choi2016} and C3K spectra. The star formation history is non-parametric and binned in time, with the SFR in each bin constrained by six log-SFR ratios ($r^1_{\rm SFH}\ldots r^6_{\rm SFH}$) between adjacent bins \citep{Johnson2021}, where each ratio follows a Student-$t$ continuity prior \citep{leja2019} with mean zero, scale 0.3, and two degrees of freedom, penalizing unphysical burstiness while permitting genuine flexibility.

Dust emission follows the \cite{Draine2007} model, with the minimum radiation field energy (\texttt{duste\_umin}), the PAH mass fraction (\texttt{duste\_qpah}), and the fraction of dust mass exposed to radiation fields above that minimum (\texttt{duste\_gamma}) all free. Dust attenuation uses the two-component screen model of \cite{Charlot2000}, with separate optical depths for young (\texttt{dust1}) and old (\texttt{dust2}) stellar populations and a variable power-law slope (\texttt{dust\_index}) for the \cite{Calzetti2000} attenuation curve.

The AGN component uses the clumpy torus templates of \cite{Nenkova2008a, Nenkova2008}, parametrized by the AGN fraction relative to bolometric stellar luminosity (\texttt{fagn}) and the torus optical depth (\texttt{agn\_tau}), both given log-uniform priors spanning several decades. Finally, the AGENT parameters \DSalpha and \DSTBB are appended with broad uniform priors.

The posterior over all 18 model parameters is sampled using the \texttt{emcee} affine-invariant ensemble sampler \citep{ForemanMackey2013}\footnote{\url{https://github.com/dfm/emcee}} as implemented within \prospector. We run 256 walkers for 15,000 steps, discarding the first 2,000 as burn-in. All parameters are initialized from a maximum-likelihood starting point obtained with \texttt{scipy.optimize}. The \DSTBB parameter is initialized at 300 while \DSalpha is initialized at 0.01.

We verify convergence in three ways. First, the integrated autocorrelation times of the AGENT and AGN (i.e., the parameters to which share the strongest degeneracy Dyson sphere emission; see \S~\ref{sec:agndegen}) parameters are typically $\sim1{,}200$--$1{,}500$ steps, so the post-burn-in chains retain effective sample sizes of a few thousand per parameter, and the posterior medians are stable between the first and second halves of the post-burn-in chains. Second, the injection-recovery campaign of \S~\ref{sec:injectRecov} provides an end-to-end validation of the inference, because the posterior medians and credible intervals it tests are exactly the quantities we report, and it recovers the injected covering fractions without bias and with the expected coverage.

Third, as a final confirmation, we refit a subset of galaxies spanning the morphological range, including all sources with strongly degenerate AGN hosts, with the \texttt{dynesty} nested sampler \citep{Speagle2020}. Wherever the photometry constrains a parameter, the two samplers agree within ${\sim}1\sigma$. For M77 (see \S~\ref{sec:resolved}), whose AGN-swarm posterior is the most strongly degenerate with our AGENT parameters, we quote the nested-sampling refit in \S~\ref{sec:resolved}. These \texttt{dynesty} test runs also illuminate a prior-dominated regime of the model itself. When the covering fraction is consistent with zero the swarm temperature is unconstrained, and below $T_{\rm BB} \sim 25$K a swarm emits outside every fitted band, so an arbitrarily large cold covering fraction trades exactly against a proportionally larger stellar mass without changing any observable. Because this cold branch is unfalsifiable with our fitted photometry, the covering-fraction constraints we report apply only to swarms warmer than $100$K. Our model-selection statistic is, in any case, insensitive to sampler convergence, because $\hat{\mathcal{L}}$ is obtained by direct optimization as described below.

To assess whether the AGENT model provides a statistically significant improvement over the standard galaxy model, we compute the Bayesian Information Criterion (BIC; \citealt{Schwarz1978}) for each fit,

\begin{equation}
    {\rm BIC} = k\ln n - 2\ln\hat{\mathcal{L}},
\end{equation}

\begin{figure*}[!tp]
    \centering
    \includegraphics[width=\textwidth]{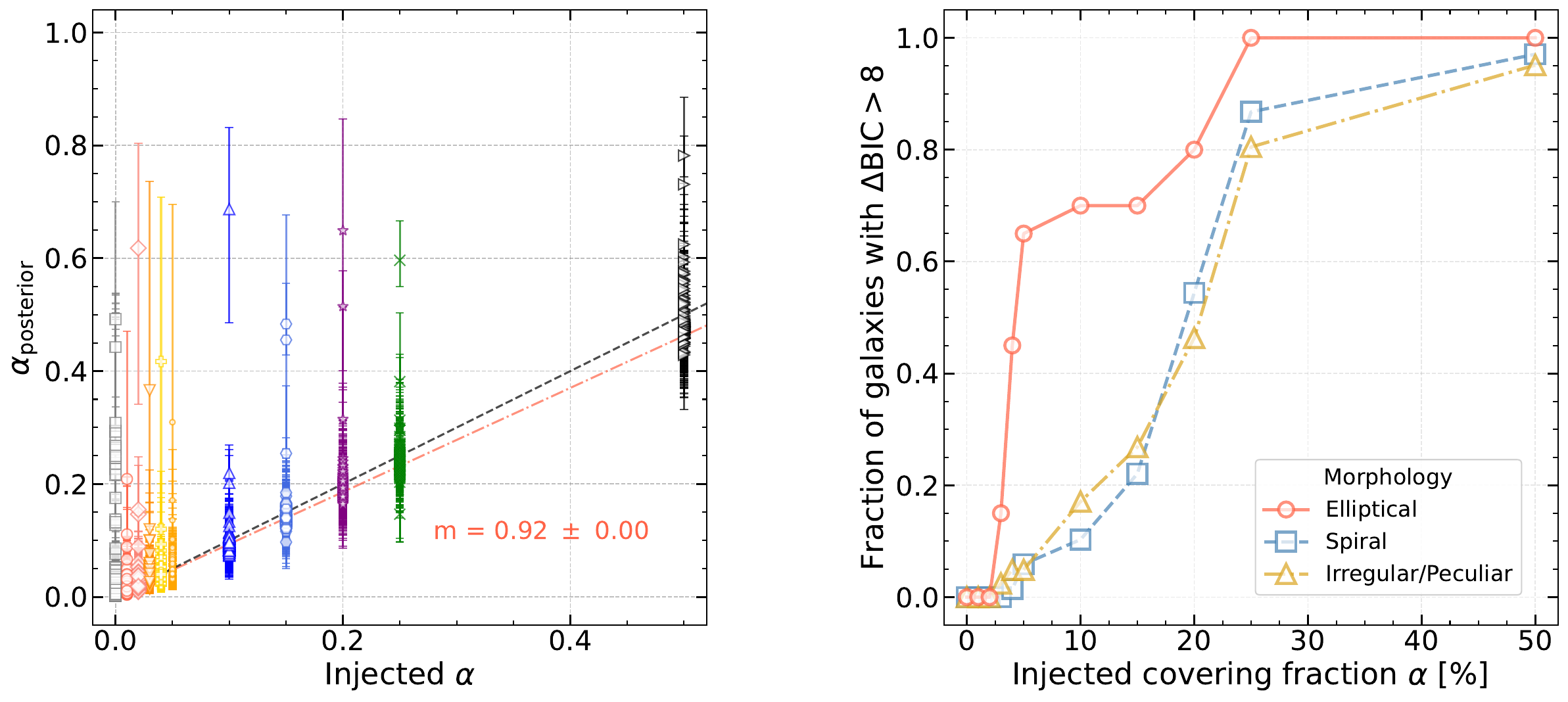}
    \caption{\textit{Left:} Injected Dyson sphere covering fraction, $\alpha$, vs. the posterior median, \DSalpha, recovered by our AGENT-enabled \prospector fits across all 129 galaxies and 11 injection levels ($N = 1{,}419$ SEDs). Each point is a single galaxy, errorbars show the 16--84\% credible interval, the black dashed line is the 1:1 relation, and the orange line and text show the best-fit slope $m = 0.92$. \textit{Right:} Fraction of galaxies of each morphological type for which the AGENT model is strongly preferred ($\Delta{\rm BIC} > 8$) as a function of injected covering fraction. Elliptical galaxies (red circles) reach $\sim50\%$ detection efficiency by $\sim4$--$5\%$ covering fraction while star-forming spirals (blue squares) require $\alpha \approx 20\%$. Irregular and peculiar galaxies (gold triangles) fall between the two. Together the two panels establish that the pipeline recovers injected covering fractions without bias, and that sensitivity is governed by how faint and predictable a galaxy's own IR emission is rather than by morphology itself.}
    \label{fig:DSaVsTruth}
    \label{fig:BICvsMorphology}
\end{figure*}

\begin{figure}[!ht]
    \centering
    \includegraphics[width=0.84\columnwidth]{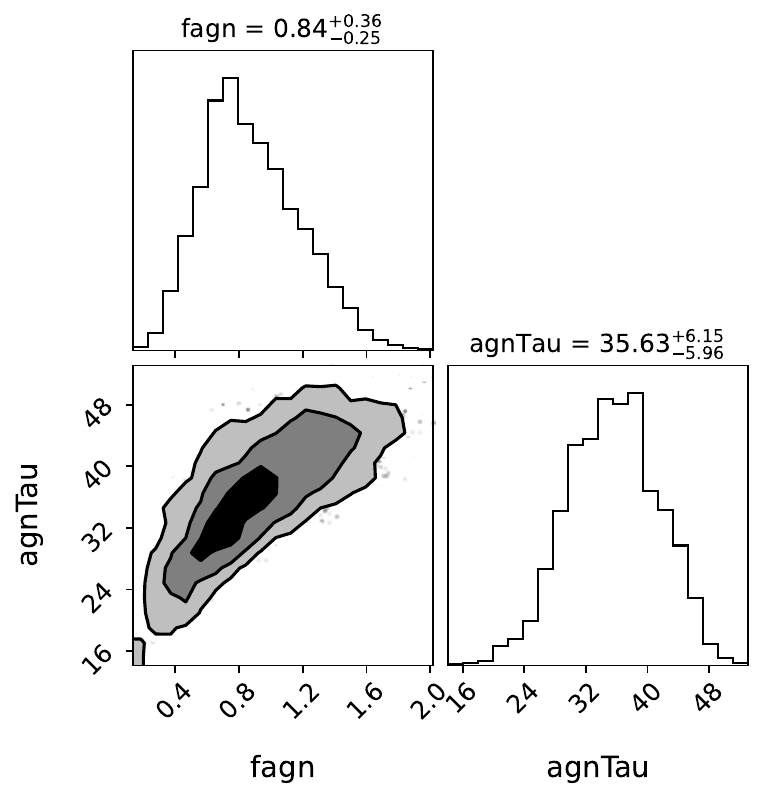}
    \\\includegraphics[width=0.88\columnwidth]{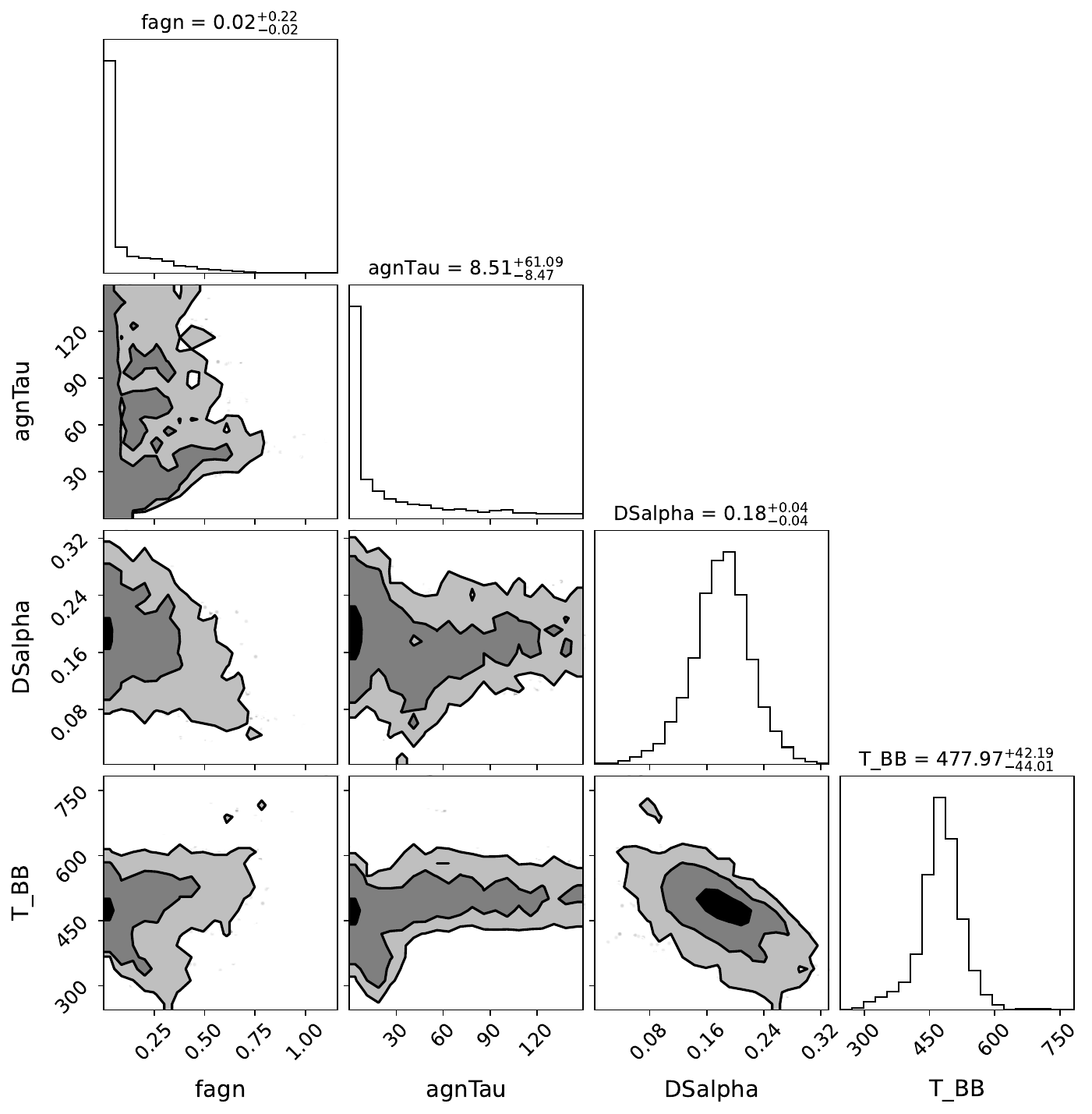}
    \caption{Posterior distributions for \texttt{fagn}, \texttt{agn\_tau}, \DSalpha, and \DSTBB for M77 (NGC~1068) when AGENT parameters are disabled (upper) and enabled (lower). This galaxy is the most extreme outlier in Figure~\ref{fig:DSaVsTruth}, where its dust-obscured AGN produces nearly featureless MIR emission that closely mimics a warm ($\sim100-300$K) blackbody, which drives \DSalpha to large values even with no injected sphere. When the AGENT parameters are disabled (upper), $f_{\rm AGN}$ rises to absorb the excess, which, for a genuinely quiescent galaxy, would imply an AGN luminosity that optical spectroscopy would either rule out or confirm, though, in M77's case, the AGN is real. When enabled (lower), the posterior transfers nearly the entire excess to the Dyson sphere component and $f_{\rm AGN}$ collapses toward zero under its log-uniform prior, yet model selection between the two configurations remains inconclusive (\S~\ref{sec:modelSelection}), so integrated photometry alone cannot distinguish the two. Spatially resolved fitting with nuclear excision can alleviate this degeneracy (\S~\ref{sec:resolved}).}
    \label{fig:cornerM77AgentOff}
\end{figure}

\noindent where $k$ is the number of free parameters, $n$ is the number of photometric bands, and $\hat{\mathcal{L}}$ is the maximum likelihood. We then compute $\Delta{\rm BIC} = {\rm BIC}_{\rm AGENT\,OFF} - {\rm BIC}_{\rm AGENT\,ON}$, taking $\Delta{\rm BIC} > 8$ as strong evidence in favor of the model containing Dyson sphere emission. Because the AGENT model adds two free parameters ($\alpha$ and $T_{\rm BB}$), the BIC penalizes these additions by $2\ln n \approx 6.4$ for our typical filter count of $n \sim 24$. A $\Delta{\rm BIC} > 8$ therefore requires a substantial improvement in maximum likelihood that cannot be attributed to overfitting.

We emphasize that $\hat{\mathcal{L}}$ is not taken directly from the posterior samples. Sampling algorithms target posterior mass rather than the maximum-likelihood point, so the highest likelihood encountered by any sampler is a noisy, run-to-run-variable estimate of $\hat{\mathcal{L}}$. We instead refine the highest-probability chain sample with a chained sequence of prior-excluded Nelder--Mead optimizations of the likelihood alone, restarting until the improvement per round falls below 0.05, and we additionally seed each member of an AGENT-on and AGNET-off pair from its partner's optimum so that neither member can stall in a local mode its partner has escaped.

The nested structure of the pair provides an internal convergence check since the AGENT-on model contains the AGENT-off model in the $\alpha \to 0$ limit, so the true optima must satisfy $\ln\hat{\mathcal{L}}_{\rm ON} \geq \ln\hat{\mathcal{L}}_{\rm OFF}$ and hence $\Delta{\rm BIC} \geq -2\ln n$, up to the small offset permitted by the $\alpha \geq 10^{-3}$ lower bound of the prior. Any deeper violation of this floor exposes an unconverged fit (e.g., \citealt{Protassov2002}). After refinement, the residual in $\Delta{\rm BIC}$ is small compared to our detection threshold of 8.

Figure~\ref{fig:ArpSpec} shows a representative \prospector fit to the star-forming galaxy Arp~256~N. The UV--optical SED and the recovered SFH both indicate a recent burst of star formation consistent with the known interacting nature of this system. The MIR to FIR emission reveals a large dust component, including prominent PAH emission features near $10^5$~\AA. The normalized residuals are all within $\sim1\sigma$, confirming that the baseline \texttt{Prospector-$\alpha$} model provides an excellent description of the data before any Dyson sphere emission is introduced. Despite having 16 to 18 parameters, these models do not overfit. Overfitting would appear as either a failed recovery of an injected signal or as a spurious detection, neither of which occurs in the $\gtrsim2,000$ fits that we perform throughout this manuscript. In \S~\ref{sec:injectRecov}, we show that we recover a nearly one-to-one relation between the injected $\alpha$ and our inferred $\alpha$. The BIC test also charges $2\ln n \approx 6.4$ for the two AGENT parameters, and no fit that uses real photometry even reaches $\Delta{\rm BIC} > 0$, so adding the 2 AGENT parameters does not cause our model to overfit.

\subsection{Resolved SED fitting and AGN excision}
\label{sec:resolvedMethod}

For a subset of nearby galaxies, we extend the injection recovery framework to the spatially resolved case. We prepare the imaging with the \texttt{piXedfit} software package \citep{abdurrouf2021}, which ingests multiwavelength data spanning the far-UV to the MIR, performs point spread function (PSF) homogenization to the resolution of the WISE W4 band (FWHM $\approx17\arcsec$), and resamples every image onto a common pixel grid. 
We then measure matched-aperture photometry in three regions, namely the full integrated light, an inner aperture that is 2 WISE W4 PSFs ($\approx34\arcsec)$ across and encloses the nucleus, and the outer region bounded by that aperture and the edge of the original catalog aperture. We fit each region separately with and without the AGENT parameters enabled. Six fits per galaxy therefore return a covering fraction and a swarm temperature for each region, together with the waste-heat-free counterpart that we compare it against.

The most persistent contaminants in any unresolved waste heat search are the dust-obscured AGNs at the centers of galaxies, which can produce nearly featureless MIR emission that closely mimics the smooth blackbody signature of a Dyson sphere. In the integrated-light case, this degeneracy is difficult to break with photometry alone. In the resolved case, however, AGN influence is concentrated in the innermost pixels of the galaxy. We can therefore excise the central region, or two times the W4 FWHM, and search for Dyson sphere emission only in the outer disk where the stellar population dominates and AGN contamination is minimal. This strategy is demonstrated in \S~\ref{sec:resolved} for the nearby Seyfert~2 galaxy M77 (NGC 1068), one of the brightest dust-obscured AGN in the local universe. While some residual AGN flux inevitably leaks into neighboring pixels through the wings of the PSF, this excision does reduce contamination in the outer spiral arms, and any candidate detection in the outer disk becomes far more credible as a result.

This analysis carries a known limitation of resolved SED fitting, which is the fact that dust responds to starlight from outside its aperture as well as to the starlight within it. Warm dust traces clustered massive stars that can lie some distance away, and diffuse cirrus emission can be powered by the general radiation field of the whole galaxy, so the energy balance that \prospector enforces within a region is only approximate. However, we argue that this limitation works in our favor for the excision test of \S~\ref{sec:resolved}. Nuclear light that escapes the inner aperture heats dust in the outer region and adds MIR emission there, and, since the torus is weakly constrained at large radii, we show blow that the swarm component will absorb its MIR excess. Nonlocal heating therefore inflates the bounds on $\alpha$ that we place on the outer region such that the decline that we measure between the integrated and outer apertures is a lower bound on the true contrast. We briefly note that a single resolved optical spectrum that measures the Balmer decrement can further alleviate this degeneracy, since dust that emits in one bin reddens the nebular lines of all light entering and leaving that bin, so a dusty region will show a high IR excess and a high decrement while a swarm will raise the IR excess and leave the decrement untouched. Although this exact test is beyond the scope of this study, we calibrate the discriminant itself against our injection suite in \S~\ref{sec:failuremodes}.


\section{Results}
\label{sec:results}

We search for Dyson sphere emission in 129 nearby galaxies spanning the full range of morphological types that are observed in the local universe. We generate 1,419 injected SEDs as described in \S~\ref{sec:galCat} and fit each twice with \prospector, once with AGENT parameters enabled and once without, for a total of 2,838 fits. The results are described below.

\begin{figure*}[!tp]
    \centering
    \includegraphics[width=\textwidth]{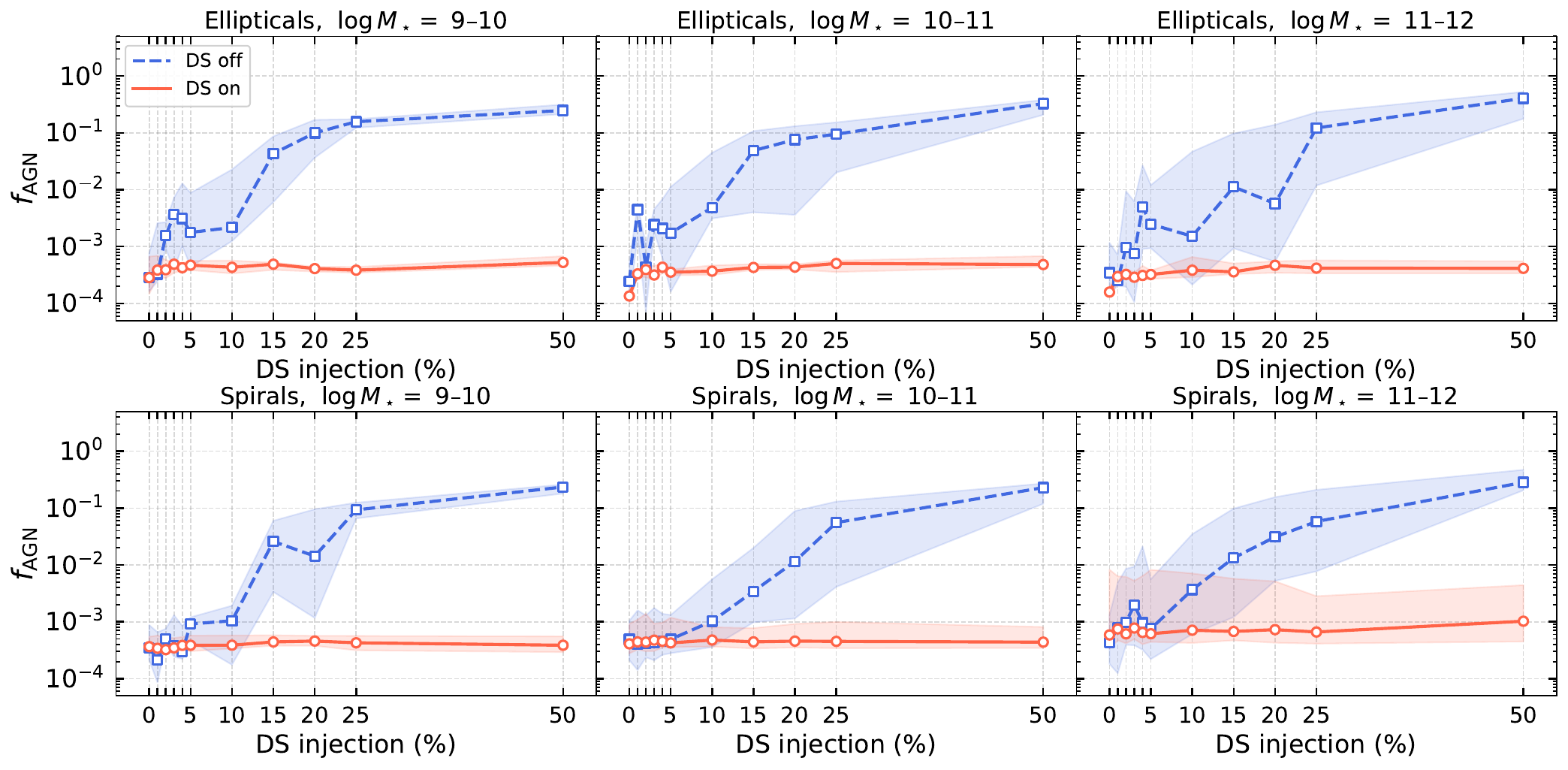}
    \caption{Population median (curves) and 16--84\% percentile band (shaded) of the inferred AGN fraction $f_{\rm AGN}$ as a function of Dyson sphere injection percentage, for elliptical (top row) and spiral (bottom row) galaxies in three stellar mass bins of $\log_{10}(M_\star/M_\odot) = 9$--$10$, $10$--$11$, and $11$--$12$ (left to right columns). The blue dashed curves with square markers show the Dyson sphere-off runs, while red solid curves with circle markers show the Dyson sphere-on runs. In the Dyson sphere-off runs, $f_{\rm AGN}$ rises nearly monotonically with injection level across all mass bins, revealing that the model is misattributing waste heat to the AGN torus. In the Dyson sphere-on runs, it remains flat with injection level, confirming that the AGENT component correctly absorbs the injected excess. The inferred torus optical depth $\tau_{\rm AGN}$ behaves in the same way and is not shown. The rising AGN fraction in the Dyson sphere-off run is therefore a diagnostic in its own right since a waste-heat-free model can incorrectly predict an AGN that the galaxy does not host.}
    \label{fig:AGNDiagnostics}
\end{figure*}

\subsection{Injection Recovery}
\label{sec:injectRecov}

Figure~\ref{fig:DSaVsTruth} (left) shows the recovered \DSalpha posteriors against the true injected values across all 129 galaxies and 11 injection levels. We recover injected Dyson sphere covering fractions with a best-fit slope of $m = 0.92$ and find that $\gtrsim95\%$ of galaxies are consistent with their injected values to within $1\sigma$. The slight underestimate relative to a 1:1 relation arises because the model allocates a small portion of the MIR budget to dust emission even when the Dyson sphere parameters are turned on.

A handful of galaxies ($\sim4$--5) consistently have their Dyson sphere covering fraction overestimated across all injection levels, including at $\alpha = 0$. In every case, these sources host a warm-to-hot dust-obscured AGN whose MIR to FIR emission closely approximates a blackbody. Figure~\ref{fig:cornerM77AgentOff} shows the posteriors on the AGN and AGENT parameters for M77 (NGC~1068), a nearby Seyfert~2 galaxy and one of the most extreme examples of this class in our sample. The current AGN model in \prospector, which is based on the \cite{Nenkova2008a, Nenkova2008} clumpy torus templates, cannot fully reproduce the emission of these objects, and the fitter compensates by attributing the residual MIR excess to Dyson sphere emission. These templates provide only a simplified description of the diverse IR spectral energy distributions of real AGN \citep{Lyu2022a, Lyu2022b}, which likely aggravates this confusion. This is the dominant systematic in our search that we address in \S~\ref{sec:resolved}, but we emphasize that in no case does an AGN generate a false positive detection with $\Delta BIC > 8$.

\subsection{Model Selection}
\label{sec:modelSelection}

Figure~\ref{fig:BICvsMorphology} (right) shows the fraction of galaxies of each morphological type for which $\Delta{\rm BIC} > 8$ as a function of injected covering fraction. For the elliptical galaxies in our sample, which are predominantly old and quiescent systems with little ongoing star formation and minimal dust, we achieve significant detections at covering fractions as low as $\sim3\%$, where $15\%$ of the ellipticals are already correctly identified as containing Dyson spheres. By $\alpha \approx 4$--$5\%$, half of the elliptical galaxies are detected. The MIR emission of these old, quiescent galaxies is dominated by circumstellar dust around AGB stars at a predictable and low level \citep{Knapp1992, Athey2002, Villaume2015}, so any excess is immediately anomalous. We emphasize that quiescence rather than morphology is what carries this sensitivity since star-forming ellipticals and quiescent disks both exist and recently quenched post-starburst systems can remain IR-bright through entirely ordinary stellar heating of residual dust. By $\alpha \approx 25\%$, every elliptical in our sample is correctly identified as containing Dyson spheres.

The picture is more complex for star-forming spiral galaxies, whose higher MIR background from ongoing star formation and interstellar dust means that a smaller fractional blackbody excess is harder to distinguish from natural variation. Only about a fifth of the spirals are detected at $\alpha = 15\%$, and a civilization must be utilizing $\alpha \approx 20\%$ of its galaxy's radiation energy budget before a $\Delta{\rm BIC} > 8$ detection becomes likely for half of them. Irregular and peculiar galaxies fall between these two extremes.

\begin{figure*}[!tp]
    \centering
    \includegraphics[width=\textwidth]{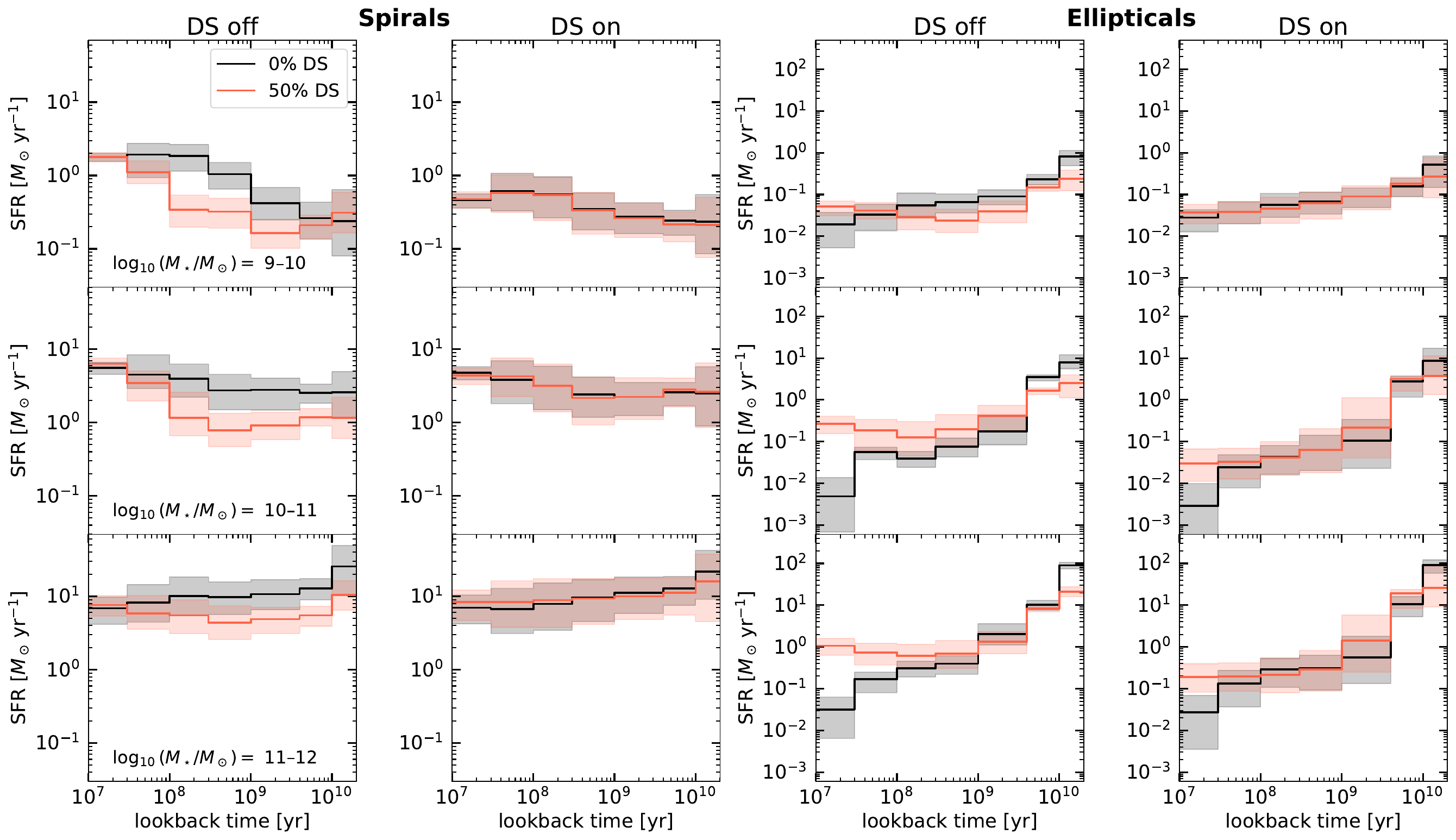}
    \caption{Population-averaged star formation histories for spiral (left two columns) and elliptical (right two columns) galaxies, each split by whether we had AGENT parameters turned off (DS-off; first and third columns) vs. turned on (DS-on; second and fourth columns). Rows correspond to three stellar mass bins: $\log_{10}(M_\star/M_\odot) = 9$--$10$ (top), $10$--$11$ (middle), and $11$--$12$ (bottom). In each panel, the black curve and gray band show the population mean and 16--84\% credible interval for 0\% DS injection, while the red curve and band show the same for 50\% injection. In the DS-off run, \prospector compensates for the unmodeled MIR excess by invoking anomalously elevated recent star formation, an effect that worsens with galaxy mass and is most dramatic for massive quiescent ellipticals. For the ellipticals, the population-median SFR in the youngest age bin rises by $1.4$--$1.8$~dex between the 0\% and 50\% cases, whereas for the spirals the rise is $\lesssim0.3$~dex and is visually washed out by the larger dynamic range of their intrinsic star formation. When AGENT parameters are turned on, the 0\% and 50\% SFHs are statistically consistent across most mass bins and morphological types, confirming that the AGENT component correctly absorbs the injected excess without biasing the inferred star formation history.}
    \label{fig:SFHgrid}
\end{figure*}

Across all morphological types, these detection limits represent a factor of $\sim3$ improvement for star-forming spirals and more than an order of magnitude for quiescent galaxies over the color-cut-based upper limits that were set by \cite{Annis1999} and \cite{Griffith2015}, which placed limits only at the $\sim50$--$80\%$ level. The contemporaneous WISE/CatWISE study of \cite{Huang2026} quotes nominally similar caps of $\alpha \lesssim 1.7$--$2.9\%$ over $T = 150$--$600$K, but the two numbers are not comparable. Their caps permit the entire observed W3/W4 flux to be waste heat and divide the sample-median luminosity cap by a fiducial Milky-Way stellar luminosity. The same accounting applied to a dusty starburst or AGN host, the galaxies that dominate our error budget, yields caps of tens of percent, and such galaxies are masked out of their sample entirely.

Our limits are instead inferred individually for every galaxy, AGN hosts included, marginalized over each galaxy's own stellar, dust, and AGN components. This is also why our per-galaxy limits are not arbitrarily tighter despite the far greater information content of the full SED since that information is already being used to constrain the dust emission that a MIR-only cap must either ignore, yielding artificially tight limits, or mask away, yielding no limit at all. Our rigorous modeling thus turns an exercise of deriving upper-limits into a per-galaxy search that yields calibrated credible intervals and a model-selection statistic for every galaxy that we can use to individually rank candidates by how anomalously IR-bright they are (see \S~\ref{sec:confirming}).

At their true, uninjected photometry, the AGENT-on fits place a per-galaxy upper limit on any genuine Dyson sphere covering fraction, which we take throughout as the $95$th percentile of the marginal \DSalpha\ posterior and write as $\alpha_{95}$. These limits span a wide range that is set by the waste-heat vs. AGN degeneracy that we discussed above. The quiescent early types, whose own IR emission is the faintest and the most predictable, carry the tightest constraints, with a median $\alpha_{95}$ of $0.3\%$ across the 29 ellipticals and lenticulars that do not host a dominant AGN, and $90\%$ of them have $\alpha_{95} \leq3\%$. Approximately half of the complete sample reaches $\alpha_{95} \leq 5\%$, whereas the dust-obscured AGN hosts that drive the injection-recovery outliers are limited only to $\alpha \lesssim 0.5$--$0.7$.

These numbers come with some caveats. First, the covering-fraction prior is uniform from $\alpha = 10^{-3}$ (Table~\ref{tab:FSPS}), so an $\alpha_{95}$ of $0.2$--$0.3\%$ lies within a couple of orders of magnitude of that floor and should be read as conservative since a prior extending to smaller $\alpha$ would move it downward rather than upward. The upper limits are also tighter than the covering fractions that our $\Delta \rm{BIC}$ model selection tests produce since the two statistics answer different questions. That is, an upper limit on $\alpha$ asks how much waste heat the photometry permits once the stellar, dust, and AGN components are marginalized over, while $\Delta{\rm BIC} > 8$ further requires that swarms improve the maximum likelihood enough to outweigh the cost of adding two additional parameters to our model. We therefore quote the injection-calibrated thresholds of Figure~\ref{fig:BICvsMorphology}, and not these upper limits, whenever we describe the sensitivity of the search. 

Still, we can convert our per-galaxy $\alpha_{95}$ limits into a population-level constraint, albeit with the caveat that our ``population'' size is $N=129$, as we can derive a per-galaxy detectability level at every covering fraction directly. That is, if a fraction, $f$, of galaxies host $300$K swarms that intercept a given share, $\alpha$, of their starlight, then the expected number of detections among our 129 real fits would be $f$ multiplied by the number of injected recoveries in that covering fraction such that the absence of detections on the real photometry excludes $f \geq 1 - 0.05^{1/N_{\rm det}}$ at $95\%$ confidence, where $N_{\rm det} = 19$, $28$, and $112$ is the number of injections recovered at $\alpha = 5$, $10$, and $25\%$. From our fits, we bound the fraction of such hosts to $f < 14.6\%$, $10.1\%$, and $2.6\%$, respectively. We note that these bounds only apply to similar surveys that target our same wavelength coverage and injected Dyson sphere temperatures (see, e.g., \S~\ref{sec:limitedFilters} and Appendix~\ref{sec:Tappendix} to see how these limits depend on wavelength coverage and swarm temperature, respectively). When applied to a survey-scale catalog, they are likely to tighten by several orders of magnitude due to the sheer sample size increase of a large-scale survey (see \S~\ref{sec:scaling}).

\subsection{Effects on AGN Parameters}
\label{sec:AGNParams}

Figure~\ref{fig:AGNDiagnostics} shows the population-median inferred AGN fraction ($f_{\rm AGN}$) as a function of Dyson sphere injection percentage, for elliptical (top row) and spiral (bottom row) galaxies in three stellar mass bins. In the DS-off run, $f_{\rm AGN}$ and the torus optical depth $\tau_{\rm AGN}$ (not shown) both rise monotonically with injection level across both morphological types and all mass bins. As more MIR flux is injected into the SED, the model, lacking the AGENT parameters to fit it, invokes a brighter and more optically thick AGN torus. The effect is most striking for elliptical galaxies, where the population-median $f_{\rm AGN}$ climbs by roughly three orders of magnitude between the 0\% and 50\% injection cases.

An analyst fitting these SEDs without a waste heat component would thus conclude that a large fraction of the elliptical population hosts luminous AGN, a conclusion that can be tested by optical spectroscopy through the absence of broad Balmer emission lines or of the narrow-line ratios that characterize genuine AGN. We caution, however, that this falsification is not always complete on its own since deep multiwavelength censuses find that MIR-selected AGN are both numerous and frequently so obscured that their optical signatures are weak \citep{Zou2022}, so X-ray or radio non-detections may be needed to close the case. This behavior provides a clear real-world diagnostic that an anomalously high inferred AGN fraction in a galaxy that does not ordinarily show other tell-tale signs of AGN emission is a red flag for unmodeled MIR emission.

In the DS-on run, both $\tau_{\rm AGN}$ and $f_{\rm AGN}$ remain approximately flat as a function of injection level, and the DS-on and DS-off medians converge near 0\% injection where neither model has an excess to explain. This confirms that the AGENT component cleanly decouples from the AGN model in the SED fit. The residual scatter in the DS-on run is largest for spiral galaxies, reflecting the genuine difficulty of separating the Dyson sphere and AGN components in galaxies with complex, dusty SEDs.

\subsection{Effects on the Star Formation History}
\label{sec:SFH}

We investigate how the inferred star formation histories change when Dyson sphere emission is present in the data but either accounted for (DS-on) or ignored (DS-off), restricting this analysis to morphologically classified elliptical and spiral galaxies in three stellar mass bins: $\log_{10}(M_\star/M_\odot) \in [9,10]$, $[10,11]$, and $[11,12]$.

\begin{figure*}[!tp]
    \centering
    \includegraphics[width=0.90\textwidth]{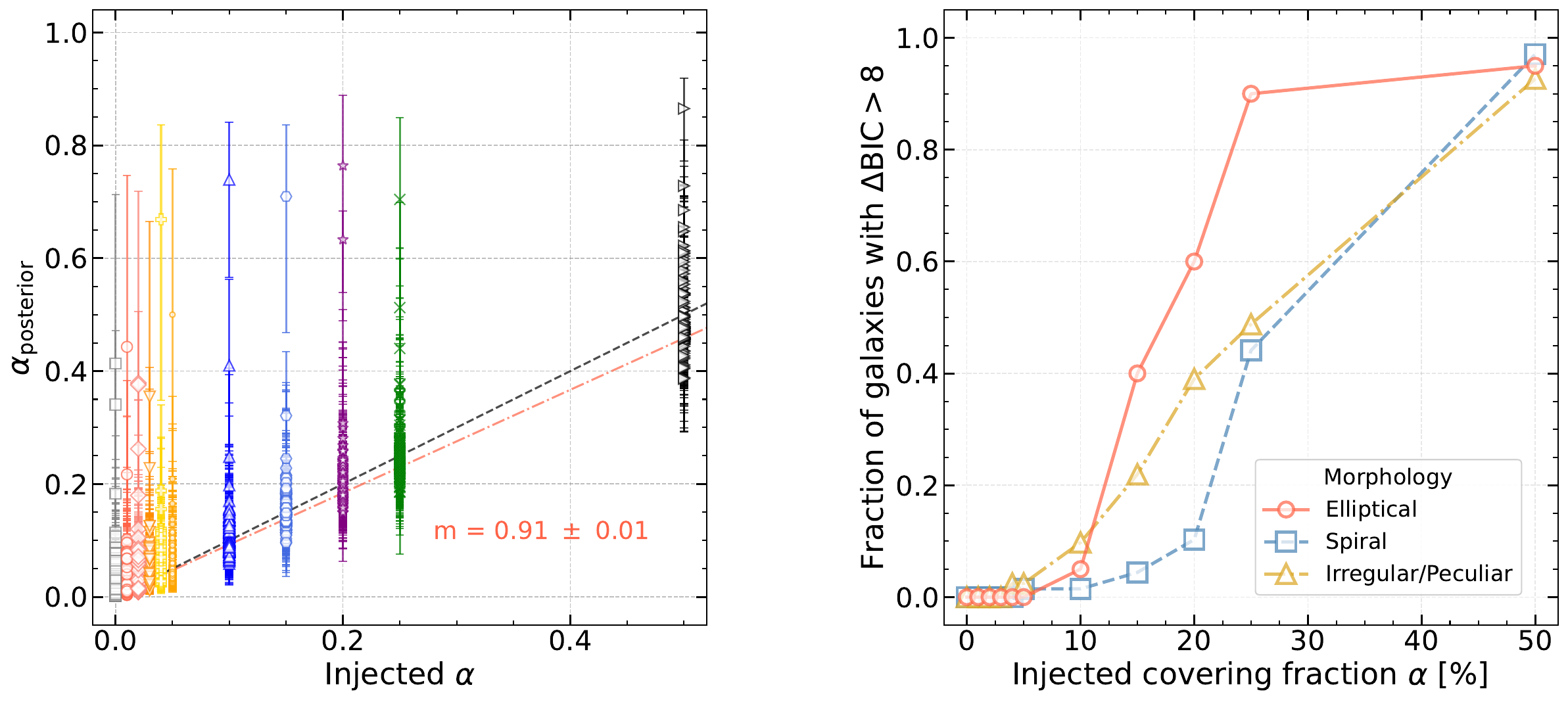}
    \caption{The same as Figure~\ref{fig:DSaVsTruth} except for the limited filter set described in \S~\ref{sec:limitedFilters}, which retains only the wide-field survey bands (GALEX FUV/NUV, SDSS $ugriz$, 2MASS $JHK_s$, and WISE W1--W4) and omits the \textit{Swift}/UVOT and \textit{Spitzer} photometry included in the fiducial set. Comparison with Figure~\ref{fig:DSaVsTruth} shows that survey-only photometry leaves the recovery slope essentially unchanged but raises the detection threshold for quiescent galaxies by a factor of ${\sim}3$--$4$, so wide-field colors alone cannot reach the covering fractions that the full filter set probes.}
    
    \label{fig:limitedFilterResults}
\end{figure*}


Figure~\ref{fig:SFHgrid} shows the population-averaged SFHs for elliptical and spiral galaxies across all three mass bins. For elliptical galaxies, the DS-off run (third column) shows a clear systematic divergence between the 0\% and 50\% SFHs (i.e., \prospector compensates for the injected MIR excess by invoking anomalously elevated recent star formation). This effect worsens with galaxy mass, and in the high-mass bin ($\log M_\star = 11$--$12$) the 50\% injection case produces recent star formation rates that are $1.6$~dex (a factor of $\sim40$) above the 0\% case. When AGENT parameters are enabled, the 0\% and 50\% SFHs are mostly consistent across most mass bins. There remains a $\sim2\sigma$ discrepancy in the most recent SFH bins where, even when AGENT parameters are enabled, \prospector still tries to increase the inferred SFHs compared to their baseline levels. However, we note that this occurs in bins of very low star formation rate ($\lesssim10^{-1}$--$10^{-2}\,M_\odot\,\rm{yr^{-1}}$) and is thus unlikely to dramatically affect the global energy balance of the galaxies.

For spiral galaxies, the picture is similar but more diffuse in that the SFH bias in the DS-off run is distributed across the SFH, dust, and AGN components rather than being solely concentrated in the recent SFH alone, reflecting the richer mix of astrophysical processes in late-type galaxies. Still, in these scenarios, the DS-on run fully recovers self-consistent SFHs across all mass bins.

\subsection{Limited Filter Set}
\label{sec:limitedFilters}


To assess sensitivity under conditions representative of an all-sky survey, we repeat all fits using only the GALEX FUV/NUV, SDSS $ugriz$, 2MASS $JHK_s$, and WISE W1--W4 bands, omitting the Swift UVOT and \textit{Spitzer} IRAC photometry. This reduces the filter count from $n \approx 24$ to $n = 14$ and removes the MIR spectral leverage that the \textit{Spitzer} [3.6]--[24] bands provide. Because the \textit{Spitzer} bands straddle the peak of Dyson sphere blackbody emission at $T \sim 300$K, their removal particularly degrades the ability to constrain the amplitude and temperature of the waste heat component.

Figure~\ref{fig:limitedFilterResults} shows the injection recovery and morphological detection results for this reduced filter set. The injection recovery trend (left panel) recovers a best-fit slope of $m = 0.91 \pm 0.01$, essentially unchanged from the full-filter result, confirming that the reduced filter set introduces no systematic bias in the mean recovered \DSalpha. However, the individual posteriors are substantially broader at all injection levels, and at injected $\alpha = 0$, recovered posteriors span up to $\alpha_{\rm posterior} \approx 0.4$--$0.7$ in a handful of galaxies, compared to a much tighter distribution in the full-filter case. The dominant high-$\alpha$ outliers are the same dust-obscured AGN hosts identified in the full-filter analysis (Figure~\ref{fig:DSaVsTruth}), consistent with the AGN-Dyson sphere degeneracy worsening as the available spectral leverage decreases.

\begin{figure*}[!tp]
    \centering
    \includegraphics[width=0.82\textwidth]{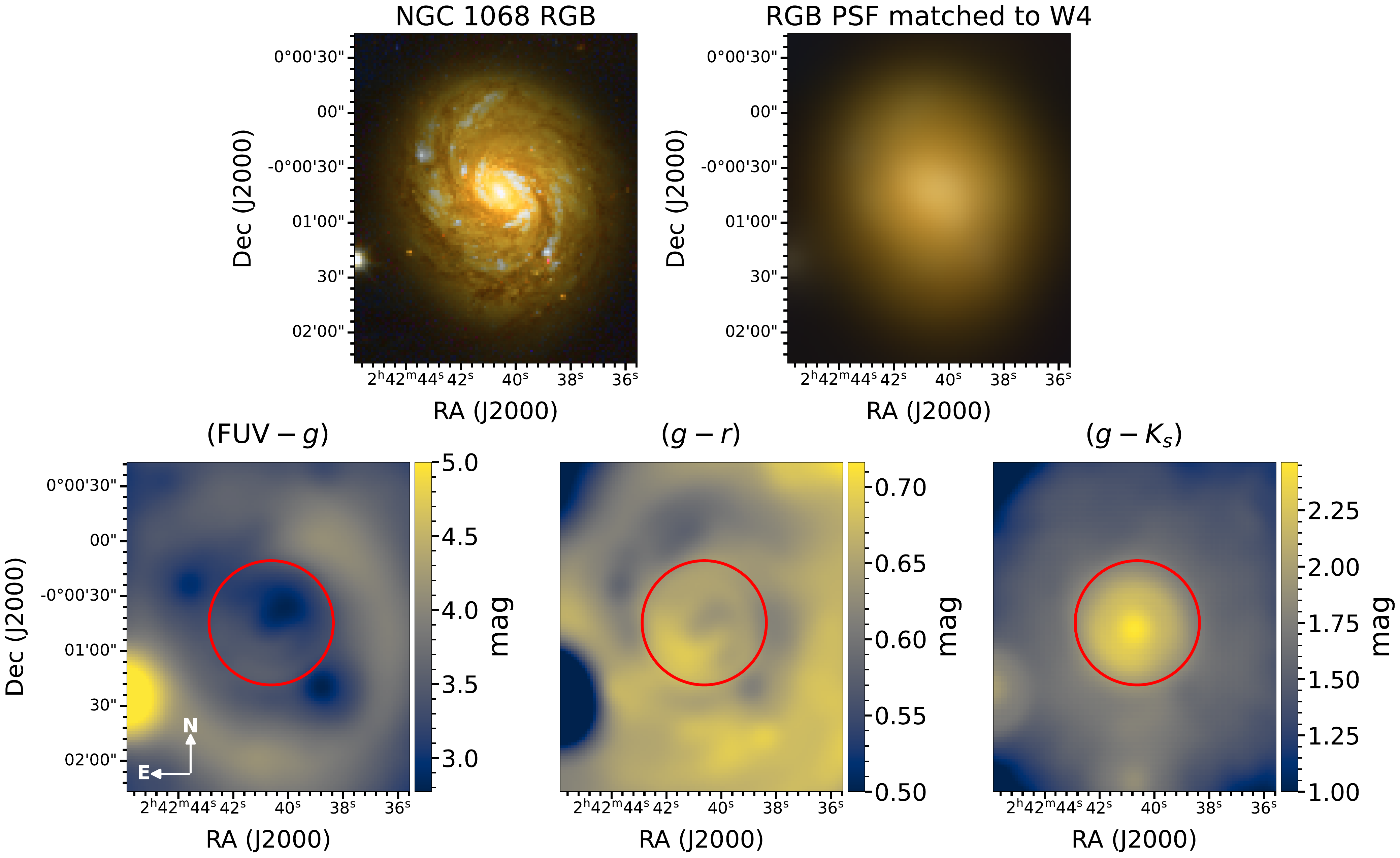}
    \caption{Resolved imaging of M77 (NGC~1068) within the \cite{Brown2014} aperture. \textbf{Top row:} a three-color rendering built from the SDSS $r$, $g$, and $u$ images, shown at the native SDSS resolution (left) and after convolution to the WISE W4 PSF (right). At the SDSS resolution, the barred spiral, its star-forming arms, and a foreground star to the southeast are all shown. At the W4 resolution, the same light collapses into a smooth, centrally concentrated ellipsoid that is only a few PSF widths across. This is the resolution at which the resolved fits and the color maps are performed below. \textbf{Bottom row:} color maps of FUV$-g$ (left), $g-r$ (middle), and $g-K_s$ (right), in AB magnitudes, constructed at the WISE W4 resolution. The red circle marks the $34\arcsec$ excision radius adopted in \S~\ref{sec:resolved}. The $g-r$ disk is comparatively featureless. The $g-K_s$ map shows the smooth gradient of the old stellar population across the disk, together with a sharp nuclear excess inside the excision radius, where AGN-heated hot dust brightens $K_s$. The nucleus is simultaneously blue in FUV$-g$ and red in $g-K_s$, which is the signature of an AGN and its circumnuclear starburst rather than of a swarm as the latter would instead dim the ultraviolet in the same region where it emits waste heat. Even at the coarse W4 resolution, this spatial contrast is what our the nuclear excision experiment exploits.}
    \label{fig:nearbymosaic}
\end{figure*}

\begin{deluxetable}{lcccc}
\footnotesize
\tablecaption{Posterior medians and 16--84\% credible intervals for the AGN and Dyson sphere parameters from our six \prospector fits to M77 (NGC~1068; i.e., the full integrated aperture as well as the inner and outer regions (see text), each run with and without AGENT parameters). Entries marked ``--'' indicate parameters that were excluded from that fit. The integrated rows quote the nested-sampling confirmation refits described in \S~\ref{sec:bayes} since this galaxy is one of the most strongly degenerate AGN-swarm hosts in our sample. The AGN parameters $f_{\rm AGN}$ and $\tau_{\rm AGN}$ carry the broad log-uniform priors of Table~\ref{tab:FSPS}, which place most of their prior weight at low AGN contributions. Because these parameters are only weakly constrained where the AGN and waste-heat components are degenerate, their posteriors and the recovered \DSalpha\ retain some sensitivity to that prior choice.}
\label{tab:resolved}
\tablehead{
  \colhead{Configuration} &
  \colhead{$f_{\rm AGN}$} &
  \colhead{$\tau_{\rm AGN}$} &
  \colhead{\DSalpha} &
  \colhead{$T_{\rm BB}$ [K]}
}
\startdata
Integrated, AGENT off & $0.84_{-0.25}^{+0.36}$   & $35.6_{-6.0}^{+6.1}$   & --                        & -- \\
Integrated, AGENT on  & $0.02_{-0.02}^{+0.22}$   & $8.5_{-8.5}^{+61.1}$  & $0.182_{-0.041}^{+0.037}$ & $478_{-44}^{+42}$ \\
Interior, AGENT off   & $0.57_{-0.21}^{+0.28}$    & $55.8_{-13.6}^{+10.5}$   & --                        & -- \\
Interior, AGENT on    & $0.000_{-0.000}^{+0.007}$ & $2.7_{-2.7}^{+46.7}$     & $0.235_{-0.079}^{+0.147}$ & $363_{-33}^{+42}$ \\
Outer, AGENT off      & $0.062_{-0.030}^{+0.100}$ & $25.1_{-25.0}^{+34.1}$   & --                        & -- \\
Outer, AGENT on       & $0.001_{-0.000}^{+0.011}$ & $0.9_{-0.9}^{+51.7}$     & $0.100_{-0.054}^{+0.137}$ & $367_{-59}^{+106}$ \\
\enddata
\end{deluxetable}

The detection thresholds for the limited filter set are substantially higher than those of the full broadband analysis. No morphological type achieves significant detections below $\alpha = 15\%$. At this level, ellipticals jump to $\sim40\%$ significant detection efficiency while Type~{\sc iii} civilizations in spirals are detected significantly $\sim4\%$ of the time, a morphological contrast of an order of magnitude. Ellipticals reach $\sim90\%$ detection efficiency by $\alpha \approx 25\%$, while spirals only reach $\sim44\%$ at the same level. Compared to the full filter set, these thresholds are $\sim3$--$4\times$ higher for ellipticals ($15\%$ versus $\sim4$--$5\%$) and moderately higher for spirals ($\sim25$--$30\%$ versus $\sim20\%$). The UV-through-MIR photometry assembled for this search is therefore a critical factor in achieving the limits reported in \S~\ref{sec:modelSelection}, and a large-scale survey restricted to GALEX, SDSS, 2MASS, and WISE photometry should be expected to yield degraded sensitivity relative to our fiducial limits.

\subsection{The Resolved Case: Masking AGN Emission}
\label{sec:resolved}

Nuclear excision reduces the AGN contribution to our recovered covering fraction. Removing the central $34\arcsec$ of M77 lowers it from $\alpha = 0.182$ in the integrated aperture to $\alpha = 0.100_{-0.054}^{+0.137}$ in the outer disk,
where the ${\sim}5\%$ of the AGN light that leaks past the 2 PSF excision radius accounts for the small residual. The contrast between the integrated and outer-region posteriors directly quantifies how much of the apparent Dyson sphere signal in integrated-light analyses of AGN hosts is attributable to nuclear emission rather than to a diffuse, civilization-scale excess. 

M77 (NGC~1068) is a nearby ($d \approx 14$~Mpc) Seyfert~2 galaxy that is one of the most extreme outliers in our injection recovery tests. Its dust-obscured AGN produces spectrally smooth, warm MIR emission that is nearly degenerate with Dyson sphere waste heat in broadband photometry, driving \DSalpha to large values even at zero injected covering fraction. The AGN's dust heating is strongest toward the nucleus, although it can extend to host-galaxy scales at a low level \citep{Symeonidis2016, McKinney2021}, whereas any genuine galaxy-spanning civilization would produce emission tracing the stellar disk \citep{Wright2014a}. Spatially separating the two regions should therefore suppress, though not fully eliminate, the AGN contribution and break the degeneracy. 

Figure~\ref{fig:nearbymosaic} shows the geometry of this test. The top row renders M77 as a single three-color image, first at the native SDSS resolution and then after convolution to the WISE W4 PSF. The spiral arms and star-forming knots that are obvious at the SDSS resolution wash out completely at the W4 resolution, where the galaxy spans only a few PSF widths. The excision must therefore operate on a coarse $34\arcsec$ aperture rather than on individually resolved structures. The bottom row shows the corresponding color maps, with the red circle marking the excision radius. The $g-K_s$ map confirms that the nuclear excess is confined inside that radius, where AGN-heated hot dust brightens $K_s$. The outer disk instead retains the smooth color gradient of its old stellar population.

We perform 4 additional \prospector fits to M77 with matched-aperture photometry across 13 bands (GALEX FUV/NUV, SDSS $ugriz$, 2MASS $JHK_s$, and WISE W1, W2, and W4). M77 saturates WISE W3, so that band is omitted from all fits of this galaxy. That is, in addition to the full integrated light of the entire galaxy, we also perform separate fits of the inner $34\arcsec$, roughly three times the homogenized W4 FWHM, and the remaining outer region of the galaxy that is bounded by the interior region and the original rectangular aperture used by \cite{Brown2014}. Then, in each of these cases, we perform further tests with AGENT parameters turned on and off. Throughout this section, we quote the nested-sampling confirmation refits that we describe in \S~\ref{sec:bayes}. If excision successfully isolates the AGN contribution, then we expect the outer-disk posteriors to show lower \DSalpha and a less extreme torus compared to either the integrated or interior fits.

In the posterior SED fits with AGENT disabled, the MIR flux decreases systematically from the integrated to the outer aperture, reflecting the declining AGN-heated dust contribution at larger radii, and the WISE W4 flux drops markedly between the interior and outer regions. The UV-through-optical portions of the SED instead remain similar across all three apertures, as expected for a galaxy whose stellar population extends across the disk.

Table~\ref{tab:resolved} lists posteriors for all six fits. In the integrated AGENT-on fit, \DSalpha rises to $0.182_{-0.041}^{+0.037}$ at $T_{\rm BB} = 478_{-44}^{+42}$K while $f_{\rm AGN}$ collapses to $0.02_{-0.02}^{+0.22}$, the log-uniform AGN prior thus cedes nearly the entire MIR excess to the AGENT parameters. When the AGENT parameters are disabled, the fitter instead prefers the extremely dusty torus that we expect of NGC~1068, reaching $f_{\rm AGN} = 0.84_{-0.25}^{+0.36}$ and $\tau_{\rm AGN} = 35.6_{-6.0}^{+6.1}$. Excising the nucleus reduces the effect and the outer region returns $\alpha = 0.100_{-0.054}^{+0.137}$, roughly half the integrated value, where $\tau_{\rm AGN}$ drops from $25.1_{-25.0}^{+34.1}$ (AGENT off) to $0.9_{-0.9}^{+51.7}$ (AGENT on). The interior region, which still encloses the nucleus, instead returns the largest covering fraction of the three apertures, $\alpha = 0.235_{-0.079}^{+0.147}$, confirming that the apparent signal is concentrated at the center of the galaxy.

We caution that $f_{\rm AGN}$ and $\tau_{\rm AGN}$ are only weakly constrained in these fits and that they carry broad log-uniform priors (Table~\ref{tab:FSPS}) that are weighted toward low AGN contributions, so, in regions where the two components remain degenerate, this prior can shift MIR flux from the AGN components onto the Dyson sphere components, biasing \DSalpha\ upward. We have shown that nuclear excision reduces but does not fully remove this sensitivity. To test this directly, we recompute all six posteriors under a linear-uniform prior on $f_{\rm AGN}$ (flat from $0$ to $3$) in place of the fiducial log-uniform prior. Because only the prior changes, the new posteriors follow from importance reweighting of the fiducial chains. This is valid here because the fiducial posteriors densely sample the entire region where the reweighted posteriors carry mass, with reweighted effective sample sizes in the thousands for the integrated pair, and a direct refit of the integrated pair under the new prior returns values consistent within ${\sim}1\sigma$.

The integrated fit is the most sensitive to the prior, where $f_{\rm AGN}$ rises to $0.31_{-0.18}^{+0.22}$ and \DSalpha\ falls from $0.182_{-0.041}^{+0.037}$ to $0.16_{-0.06}^{+0.06}$, so the fiducial log-uniform prior does inflate the integrated covering fraction where the AGN and the swarm are degenerate. The outer-disk result is far less sensitive, moving from $\alpha = 0.100_{-0.054}^{+0.137}$ to $\alpha = 0.15_{-0.07}^{+0.33}$ and remaining consistent with zero at ${\approx}2\sigma$, so the resolved conclusions of this section are not driven by the choice of AGN prior.


\section{Discussion}
\label{sec:discussion}

\subsection{Principal Results}
\label{sec:principal}

We present the most robust SPS-based search for galaxy-spanning Dyson sphere emission, where we have incorporated the AGENT formalism into \texttt{FSPS} and \prospector, which we used to perform injection recovery tests on the \cite{Brown2014} galaxy atlas. Our pipeline successfully recovers injected Dyson sphere covering fractions with a best-fit slope of $m = 0.92$, where we show that $\gtrsim95\%$ of the covering fractions are consistent with their injected values to within $1\sigma$. By injecting waste heat at the SSP level and sampling the full posterior, we produce not only point estimates but calibrated credible intervals for every parameter for individual galaxies, completely bypassing the need for population-level color cuts. However, we note that these upper limits only apply to our searches using our exact photometric filters, and can thus only improve or degrade as more or fewer photometric bands are used (e.g., \S~\ref{sec:limitedFilters}). 

This is a fundamental advance over previous searches (e.g., \citealt{Annis1999, Griffith2015}), which relied on color cuts applied to integrated photometry and were limited to upper limits of $\sim50$--$80\%$ of a galaxy's starlight. More recently, \cite{Huang2026} quoted nominal caps of $\alpha \lesssim 1.7$--$2.9\%$ from WISE flux limits alone, though these assume a fiducial stellar luminosity and they do not subtract host emission or exclude AGN- and starburst-like galaxies from their sample (\S~\ref{sec:modelSelection}). Our approach detects civilizations using as little as $\sim4$--$5\%$ of their host galaxy's starlight in elliptical galaxies, while accounting for the entire energy budget of the galaxy in a principled way. Star-forming spirals, whose higher MIR background from dust and ongoing star formation makes any excess harder to distinguish, require covering fractions $\gtrsim20\%$ for a robust detection. Crucially, our Bayesian approach provides a rigorous detection criterion that naturally prioritizes more parsimonious models, something unattainable with heuristic color cuts.

The sensitivity hierarchy is physically intuitive, and it tracks quiescence more fundamentally than morphology. Old, quiescent galaxies (i.e., the bulk of our morphologically classified ellipticals) have simple, well-characterized SEDs. That is, their MIR emission is normally dominated by circumstellar dust around AGB stars at a predictable and low level \citep{Knapp1992, Athey2002, Villaume2015} or by centrally located AGN, so any excess is immediately anomalous. Star-forming spirals carry a richer and more variable mix of PAH emission and warm interstellar dust, both of which compete with the waste heat signal. This makes old, quiescent galaxies, specifically their outer edges, the optimal targets for any extragalactic technosignature survey. 

Indeed, our resolved search can be extended to very nearby galaxies, where the effective number of stars per resolution element can be of order $\sim10^2-10^6$ (e.g., \citealt{conroy2016}; \citealt{cook2019}; \citealt{cook2020}). For these quiescent galaxies, with sensitivities on \texttt{DSalpha} as low as $\sim3\%$, this means that we can potentially identify Type~{\sc iii} civilizations that are harvesting energy from just a few dozen to a few thousand stars. To show this with a back-of-the-envelope estimate, consider the fact that one $17\arcsec$ WISE W4 Atlas beam \citep{Cutri2013} subtends $\sim330 \, \rm{arcsec}^2$, which is $\sim20\,\rm{pc}^2$ at the Large Magellanic Cloud (LMC; \citealt{vanderMarel2001, vanderMarel2002}) and $\sim 5\times10^3\,\rm{pc}^2$ at M31 \citep{Tamm2012}. A stellar population supplies $\sim2-3$ living stars per solar mass for a \citet{Kroupa2001} IMF, so the disk of the LMC, which carries $\sim2.7\times10^9M_\odot$ on a $1.3\,\rm{kpc}$ exponential scale length \citep{vanderMarel2002}, thus has $\sim10^4-10^6$ stars per beam across its inner three scale lengths. Meanwhile, the disk of M31, with its mass of $5.6\times10^{10}M_\odot$ \citep{Tamm2012}, has $\sim10^4-10^6$ stars per beam, while typical dwarf spheroidals contain $\sim10^3$ stars per beam \citep{McConnachie2012}. These populations sit in the semi-resolved regime of $\sim10^1-10^6$ stars per resolution element for which pixel color-magnitude modeling was developed \citep{conroy2016, cook2019, cook2020}. 

Given the sensitivities of our injection tests, this calculation provides a very strong case for resolved SED fitting of nearby galaxies as a means to detect Dyson spheres. We caution, however, that such searches are unlikely to catch a galaxy in the act of being settled. \cite{Carrigan2010, Carrigan2012} describe ``Fermi bubbles,'' or growing voids in the visible light of a galaxy that expand as a society cloaks more stars. Even settlers traveling at $10^{-3}$ to $10^{-2}$ times the speed of light would span a galaxy in only $\sim1-100$ million years, and settlement simulations also show that even conservative ship ranges and launch rates leave a galaxy endemic with technology on similar timescales \citep{Carroll-Nellenback2019, Wright2021}. Since the transition period is brief, essentially every galaxy should either be untouched or fully settled. Resolved searches therefore target the two ends of this, revealing either a single anomalous region that reprocesses a few percent of its local starlight, or, in a galaxy already saturated with swarms, testing whether the IR excess traces the stellar disk as one would expect if the source of the IR excess is technological in origin.

Still, these numbers invite the Earth Detecting Earth exercise of \citet{Sheikh2025}, who asked at what distance present-day instruments could recover Earth's own technosignatures. Suppose that an astronomer in the LMC pointed our pipeline at the solar neighborhood such that their WISE W4-matched beam covers the same $\sim20\,\rm{pc}^2$ area around the Sun that we see when observing the LMC. Viewed face-on, the solar neighborhood has a surface density of $27\pm3\,M_\odot$~pc$^{-2}$ \citep{McKee2015} and a surface brightness of $24$--$30\,L_\odot\,\rm{pc}^{-2}$ in the $V$ and $I$ bands \citep{Flynn2006}, so this hypothetical beam holds roughly $500\,M_\odot$ in a thousand stars that emit $\sim500\,L_\odot$, of which Sirius~A alone supplies $\sim25\,L_\odot$ \citep{Bond2017}. Our injection tests require covering fractions of $\sim3$--$20\%$, or $\sim15$--$100\, L_\odot$ of the light within the hypothetical beam. These are back-of-the-envelope figures, good to a factor of a few since the surface brightnesses are band luminosities rather than bolometric ones and we assume the LMC astronomer is viewing us face-on rather than at its true inclination relative to the Milky Way's disk, but the conclusion does not depend on that precision. Still, a swarm enclosing most of Sirius~A alone could place our solar neighborhood near the detectable regime, especially if resolved JWST or Hubble Space Telescope imaging are used, while a complete Dyson sphere around the Sun falls an order of magnitude below the same threshold. The LMC astronomer only confidently detects humanity after we have cloaked the equivalent of every star within $\sim5$--$10$~pc of the Sun. \cite{Wright2023} estimates that humanity has placed $\sim0.1$~km$^2$ of solar panels around the Sun, intercepting $\sim10^{-19}\,L_\odot$ and producing $\sim40$~MW of waste heat, which is $22$ orders of magnitude below the light in our hypothetical beam. Even if every star in the beam hosted a civilization at our present scale, the combined signal would still fall a factor of $\sim10^{17}$ short of our detection threshold. Searches like the ones presented here are thus only sensitive to radiative energy capture that is $\sim20$ orders of magnitude beyond humanity's current limits.
  
\subsection{Signatures of Unmodeled Waste Heat}
\label{sec:failuremodes}

When Dyson sphere emission is present but unmodeled, the fitter displays two characteristic failure modes that are themselves useful diagnostics. First, it drives $f_{\rm AGN}$ and $\tau_{\rm AGN}$ to anomalously high values that scale monotonically with the injected covering fraction, a response that would be immediately suspicious for galaxies that do not show other canonical AGN markers. Several authors (e.g., \citealt{Penrose2002}; \citealt{hsiao2021}; \citealt{Curtis2026}; M. Elvis, in preparation) have also pointed out that AGN, or black holes in general, would be very lucrative energy sources for Type~{\sc iii} civilizations, and, in some cases, these societies might significantly affect the MIR spectrum of AGN. Future studies are needed to fully constrain, using radiative transfer simulations, how a Dyson sphere affects the dusty-torus models used in our fits (e.g., \citealt{Nenkova2008a, Nenkova2008}). Second, Dyson sphere emission inflates recent star formation rates, particularly for massive ellipticals, to values an order of magnitude above those expected for universally quiescent systems. Together, these signatures constitute a practical red-flag checklist such that an anomalously luminous AGN posterior or an nonphysically elevated recent SFH in a quiescent galaxy are both reasons to suspect unmodeled MIR emission.

\subsection{The AGN Degeneracy and Resolved Fitting}
\label{sec:agndegen}

Our dominant systematic is the degeneracy between warm AGN torus emission and Dyson sphere waste heat. The $\sim4$--5 strongly dust-obscured AGN hosts in our sample mimic the waste heat signal at all injection levels, and nuclear excision halves the covering fraction recovered for the one galaxy where we demonstrate it (\S~\ref{sec:resolved}). The natural question is how widely can this same test can be applied.

For our AGN excision method to work, a galaxy must be large enough on the sky to leave a large enough area beyond the excised nucleus. Our excision removes a fixed area of radius $r_{\rm ex} = 34\arcsec$, which is twice the effective PSF of the WISE W4 Atlas Images that have FWHM $\theta = 17\arcsec$ \citep{Cutri2013}, so a galaxy needs an isophotal radius $R$ satisfying $\pi(R^2 - r_{\rm ex}^2) \geq 1.13\,N\,\theta^2$ to retain $N$ independent resolution elements outside the excised core. Retaining $N = 25$ requires an isophotal diameter $D_{25} \gtrsim 2.0\arcmin$, which roughly $3{,}000$ galaxies in the Third Reference Catalog \citep{deVaucouleurs1991} satisfy, and retaining $N = 100$ requires $D_{25} \gtrsim 3.6\arcmin$, which roughly $700$ satisfy. Still, angular size is necessary, but not sufficient, for our AGN excision method to work. The beam count is a purely geometric bound that assumes that every resolution element outside the nucleus delivers usable photometry, but the WISE W4 surface brightness limit instead sets how far into a disk one can actually measure a MIR signal, so the outer regions of many galaxies will inevitably fall below the signal that a fit requires.
  
\subsection{Scaling Up}
\label{sec:scaling}

A natural next step is to apply this pipeline at scale to the $\sim10^7$ galaxies with archival MIR and multiwavelength photometry, which would require replacing the computationally expensive \prospector sampler with a sophisticated machine learning framework. The injection tests presented here will serve as the validation set for that future effort. Indeed, the characteristic bias patterns we identify (i.e., elevated AGN fractions and anomalous SFHs) provide concrete signatures to search for in any large-scale catalog of \prospector fits. Concretely, one could draw galaxy parameters from a population-level prior, forward-model each SED with the same AGENT-enabled \texttt{FSPS} model used here, inject empirical per-band noise, and then train a neural density estimator that returns full posteriors on $\alpha$ and $T_{\rm BB}$ in a fraction of a second rather than the CPU-days required per \prospector fit \citep{cranmer2020, Wang2023}. Indeed, L. Shi et al. (in preparation) demonstrate that this approach accelerates \prospector-quality inference to $\sim0.02$~s per galaxy, fast enough to screen every WISE-detected galaxy for the failure-mode signatures that we discussed in \S~\ref{sec:failuremodes}. Such an all-sky search will be the subject of our next manuscript. 

Survey-scale fitting also has a track record of unveiling new classes of galaxies through anomalous broadband photometry, such as when volunteers inspecting SDSS imaging isolated the Green Peas, compact galaxies whose extreme [O~III] $\lambda5007$ emission distorts their broadband colors \citep{Cardamone2009}. Galaxies that no natural configuration of our model can reproduce would be flagged by the same screening and cataloged as MIR anomalies by our pipeline. This screening also naturally guards against some of the known deficiencies of the stellar models themselves. For instance, the hot evolved stars that power the UV upturn of old populations remain uncertain in current SPS models \citep{Conroy2009, Johnson2021}, but an underpredicted upturn produces UV residuals rather than a spurious MIR excess, and the unexplained light pushes a fit away from a cloaking interpretation rather than toward one since Dyson spheres can only remove UV light, not produce it.

Lastly, a catalog of this scale could also help the radio SETI community. Every targeted radio pointing also records the background objects inside its primary beam.\citet{WlodarczykSroka2020} turned that fact into a stronger limit on the number of radio transmitters in the Galaxy by counting every Gaia star inside every Breakthrough Listen beam, growing the sample size from 1,327 stars to 288,315. \citet{Garrett2023} repeated the exercise for galaxies and found 143,024 of them inside 469 of the Breakthrough Listen fields. Their error budget to was dominated by the fact that they had no star count for each those galaxies, so they had to assume values similar to the Milky Way. A survey-scale run of the pipeline that we present in this manuscript could thus measure a stellar mass for each of those galaxies, which would provide tighter constraints on their calculations.

\begin{figure*}[!tp]
    \centering
    \includegraphics[width=0.82\textwidth]{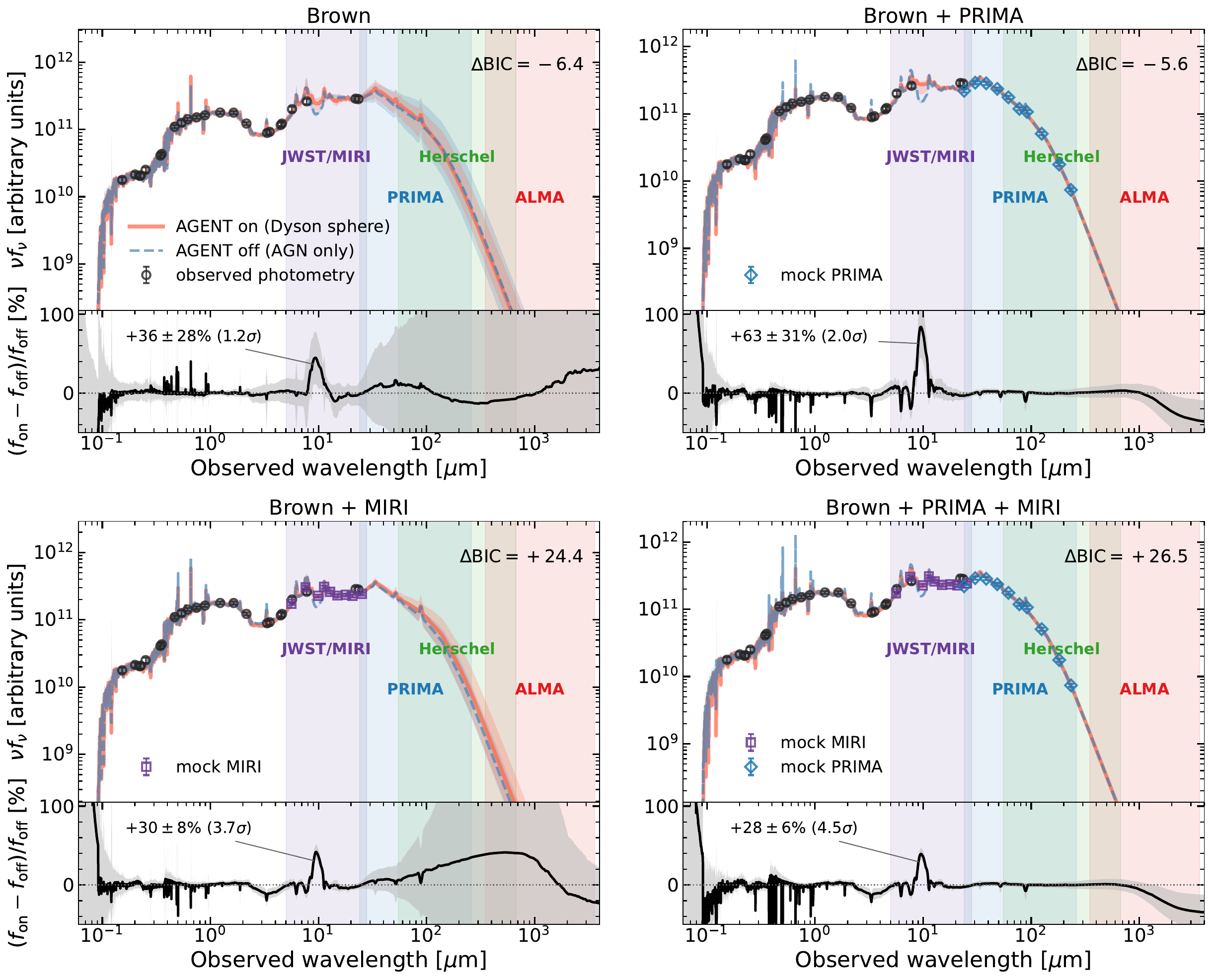}
    \caption{Where the Dyson sphere on and off fits of M77 differ most, and how new photometry changes the comparison. In each panel, the upper sub-panel shows the median posterior AGENT-on (solid red) and AGENT-off (dashed blue) model spectra with shaded $16$--$84\%$ posterior bands and the fitted photometry overplotted (open black circles), while the lower sub-panel shows the percent deviation of the AGENT-on model relative to the AGENT-off model, with the shaded band giving the $1\sigma$ uncertainty propagated in quadrature from the two independent fits. The top-left panel fits only the real \cite{Brown2014} photometry, as in the rest of this work. In the remaining panels, we append synthetic photometry generated from the best-fit AGENT-on model (blue diamonds for the ten PRIMAger bands and violet squares for the nine JWST/MIRI imager bands, both with $5\%$ uncertainties) and refit both models, quoting each panel's $\Delta{\rm BIC} = {\rm BIC}_{\rm AGENT\,off} - {\rm BIC}_{\rm AGENT\,on}$ and the significance of the deviation in the MIRI region. With the Brown photometry alone (top left), the shared predictive band swallows the entire far IR and the $10~\mu$m bump is significant at only $1.2\sigma$. Adding PRIMA (top right) pins the FIR continua of both models jointly, collapsing the band and concentrating the disagreement into the $10~\mu$m silicate and waste-heat region at $+63\pm31\%$ ($2.0\sigma$), squarely within reach of JWST/MIRI. Adding MIRI photometry tests that region directly at $3.7\sigma$ and carries the model selection to $\Delta{\rm BIC} = +24.4$ (bottom left), and the combination of both facilities (bottom right) yields $\Delta{\rm BIC} = +26.5$ with a $4.5\sigma$ deviation. Shaded vertical spans mark the wavelength coverage of JWST/MIRI, PRIMA, Herschel, and ALMA. Taken together, the four panels show that FIR photometry forces the two hypotheses into a single falsifiable configuration, which MIR imaging then adjudicates in either direction.}
    \label{fig:diffspec}
\end{figure*}

\subsection{Future Observations}
\label{sec:futureobs}

As the limited-filter test of \S~\ref{sec:limitedFilters} demonstrates, the limits derived here are ultimately set by the wavelength coverage available around the waste-heat blackbody peak. The proposed FIR observatory PRIMA \citep{Glenn2025}, with continuous imaging and spectroscopy from $24$ to $261~\mu$m, would directly sample the cold-dust spectral energy distribution that is the dominant confounder for waste heat. PRIMA spectra would thus constrain swarm temperatures below $\sim150$K that lie beyond the reach of WISE and \textit{Spitzer} IRAC imaging. Incorporating PRIMA photometry into the AGENT forward model would therefore tighten the covering-fraction limits presented here and extend them to cooler, more advanced swarms.

On the near-IR side of the waste-heat peak, the recently launched Spectro-Photometer for the History of the Universe, Epoch of Reionization, and Ices Explorer \citep[SPHEREx;][]{Dore2014, Crill2020} is conducting an all-sky spectral survey in 102 channels spanning $0.75$--$5~\mu$m, revisiting the full sky every six months and serving the data publicly. Photometry synthesized from its spectra would restore much of the leverage that the survey-only filter set of \S~\ref{sec:limitedFilters} loses relative to the atlas, anchoring the stellar continuum and the circumstellar AGB dust while sampling the wavelengths where the hot swarms of Appendix~\ref{sec:Tappendix} peak. Folding SPHEREx channels into the AGENT forward model is therefore the most immediate coverage upgrade available to a survey-scale search.

A way to ask which future data would most help is to compare our best-fit models with and without Dyson spheres for the same galaxy and see where they disagree. The top-left panel of Figure~\ref{fig:diffspec} shows the median posterior AGENT-on and AGENT-off model spectra for M77 together with the percent deviation between them. The two models are tightly constrained to agree wherever we have photometry, from the ultraviolet through the WISE and \textit{Spitzer} MIR, and their predictions separate only in the mid-to-far IR, where the deviation reaches $+36\pm28\%$ ($1.2\sigma$) near $10~\mu$m before the shared predictive band swallows the entire far IR. We emphasize that the width of that band does not mean that a FIR measurement carries no information since the band is wide precisely where the current data provide no constraint, and a new measurement acts by collapsing that shared freedom in both models jointly. The \textit{Herschel} measurements of \S~\ref{sec:confirming} demonstrate this directly, where three PACS points collapse the shared FIR freedom for NGC~7592 and measure the temperature of its residual excess, identifying it with $53$K interstellar dust.

To quantify the same point for PRIMA and JWST, we generate synthetic photometry from the best-fit AGENT-on model of M77 in the ten PRIMAger bands (the six hyperspectral channels spanning $24$--$78~\mu$m and the four polarimetric bands at $92$--$235~\mu$m, following \citealt{Ciesla2025}) and the nine JWST/MIRI imager bands, all with $5\%$ uncertainties, append each combination to the real photometry, and refit with both our AGENT on and off models (Figure~\ref{fig:diffspec}). The PRIMA points pin the FIR continua of both hypotheses jointly, collapsing the predictive band and concentrating the models' remaining disagreement into the $10~\mu$m silicate and waste-heat region, where the deviation sharpens from $+36\pm28\%$ ($1.2\sigma$) to $+63\pm31\%$ ($2.0\sigma$).

Model selection alone remains inconclusive for this extreme AGN host. When the synthetic galaxy carries M77's own AGENT-on posterior median, which places a covering fraction near $0.18$ at roughly $400$K, the refit returns $\Delta{\rm BIC} = -5.6$ while the swarm-free fit instead gives $\Delta{\rm BIC} = -6.9$. Neither case approaches $\Delta{\rm BIC} > 8$ since the free torus can still absorb a ${\sim}480$K swarm even with the far IR pinned. The role of FIR photometry is therefore not to adjudicate the degeneracy on its own but to force the two hypotheses into a single falsifiable configuration. Synthetic MIRI photometry, which samples the $10~\mu$m region directly, raises the significance of that discriminating bump to $3.7\sigma$ on its own and to $4.5\sigma$ when combined with PRIMA, with the model selection moving to $\Delta{\rm BIC} = +24.4$ and $+26.5$, respectively, while the same refits of the sphere-free synthetic observations remain null at $\Delta{\rm BIC} = -6.9$ and $-7.5$. That is, wide-field FIR photometry with PRIMA will make targeted, expensive JWST/MIRI observations decisively informative. ALMA anchors the cold Rayleigh-Jeans tail at still longer wavelengths, where the two models reconverge, which helps separate a cool swarm from genuinely cold dust.

IR spectroscopy of the PAH features is informative here in a more indirect way. While the features themselves trace star formation rather than Dyson swarms, however, Dyson spheres do intercept the photons that excite them, so a large covering fraction still suppresses PAH emission relative to the star formation rate inferred from the unobscured UV and optical light (Figure~\ref{fig:FSPSTest}), so a PAH deficit at fixed star formation rate is a useful consistency check on any candidate.

\subsection{Confirming a Candidate}
\label{sec:confirming}

Should a future application of this pipeline yield a galaxy for which our BIC test decisively favors a reprocessed Dyson sphere component over stellar, nebular, dust, and AGN emission, establishing that the source is genuinely due to Dyson spheres rather than an unusual but natural object would require a sequence of follow-up observations of increasing cost and diagnostic power. On the merit axes of \citet{Sheikh2020}, galaxy-scale waste heat scores strongly on detectability but poorly on ambiguity, so this sequence, rather than the detection statistic alone, carries the burden of proof.

\begin{figure*}[!tp]
\centering
    \includegraphics[width=0.88\textwidth]{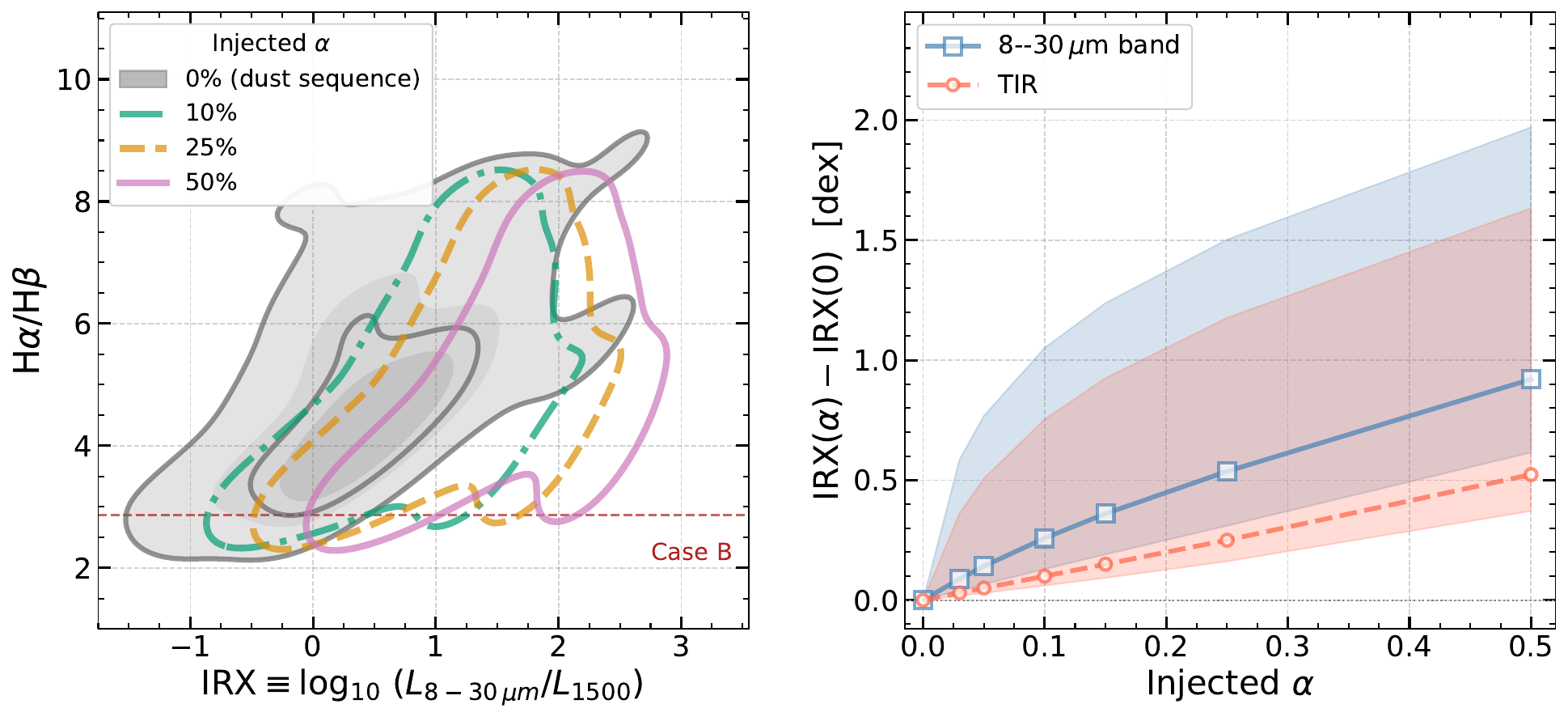}
    \caption{The IR-excess--Balmer-decrement plane under our Dyson sphere injection tests. The IR excess is ${\rm IRX} \equiv \log_{10}(L_{\rm IR}/L_{1500})$, the ratio of the luminosity integrated over a rest-frame IR band to the monochromatic $1500$~\AA\ luminosity. \textit{Left:} predicted H$\alpha$/H$\beta$ against the $8$--$30~\mu$m IR excess for the fitted models of all 129 galaxies. The gray filled contours show the uninjected population, which traces the natural dust sequence in which nebular reddening and IR excess rise together, and the colored contours show the same galaxies with swarms injected at $\alpha = 10$, $25$, and $50\%$. The dashed line marks the intrinsic dust-free Case~B recombination ratio of 2.86. \textit{Right:} the displacement of each galaxy from its own uninjected position, with points and shaded bands showing the population median and 16--84\% range for two definitions of $L_{\rm IR}$, the $8$--$30~\mu$m band (blue squares) and the total IR (TIR) band of $8$--$1000~\mu$m (red circles). The Balmer decrement is invariant under injection, so this horizontal displacement is the entire observable signal, and it is largest for the least dusty hosts. A galaxy whose IR excess is too large for its Balmer decrement therefore cannot be explained by ordinary dust, which makes this plane an inexpensive archival pre-selection for scaled-up searches.}
    \label{fig:irxbalmer}
\end{figure*}

The least expensive diagnostic is optical spectroscopy, which supplies several independent tests from a single archival spectrum. A spectrum that exhibits broad or narrow AGN emission lines, or one that yields a recombination-line-derived star formation rate well below the value implied by the IR excess, would immediately point to a conventional explanation. The Balmer decrement, especially when paired with the Paschen lines from near-IR spectroscopy, is the gold-standard measure of nebular dust attenuation, and it separates a swarm from dust because a MIR excess produced by ordinary dust must be accompanied by correspondingly high nebular reddening whereas reprocessing by a swarm requires none. A strong IR excess in a galaxy with a low Balmer decrement therefore cannot be attributed to dust. Appendix~\ref{sec:massAppendix} and Figure~\ref{fig:irxbalmer} calibrates this discriminant against our injection suite, where the dusty interpretation of a cloaked galaxy over-predicts the population-median decrement by up to a factor of $\sim1.5$.

A sufficiently deep spectrum also yields a stellar velocity dispersion and hence a dynamical mass. A swarm, however, will hide stellar light without removing stellar mass, so the stellar mass inferred from the dimmed SED of a galaxy with genuine Dyson spheres would fall systematically below the expectation from the stellar-mass--dynamical-mass relation. A candidate lying several sigma below that relation would corroborate hidden luminosity, though we regard this as a supporting rather than decisive test, since gas and dark matter also contribute to dynamical masses. Appendix~\ref{sec:massAppendix} calibrates this expectation against the same injection suite, where the hidden-mass offset exceeds twice the observed scatter of the stellar-mass--dynamical-mass relation only for covering fractions above ${\sim}40\%$. For many nearby galaxies, the required spectra already exist in public archives, so these two tests, along with the quantitative limits on each derived in Appendix~\ref{sec:massAppendix}, make the tiering of the confirmation strategy above possible for some sources.

Figure~\ref{fig:irxbalmer} quantifies the Balmer-decrement discriminant by forward modeling the atlas itself, where we show that injecting swarms into the fitted models of all 129 galaxies raises each galaxy's IR excess while leaving its nebular reddening untouched. We measure the excess as ${\rm IRX} \equiv \log_{10}(L_{\rm IR}/L_{1500})$, with $L_{\rm IR}$ the luminosity integrated over a rest-frame IR band and $L_{1500}$ the monochromatic $1500$~\AA\ luminosity \citep[e.g.,][]{Meurer1999, Kennicutt2012}. The reddening is untouched because the swarm absorbs the ionizing continuum and the recombination lines in the same proportion, so the predicted H$\alpha$/H$\beta$ is unchanged to better than $10^{-4}$~dex at any covering fraction. The uncloaked population traces the familiar dust sequence in which reddening and IR excess rise together, while a cloaked galaxy slides off that sequence horizontally. At injected $\alpha = 50\%$ the median displacement is $0.9$~dex in an $8$--$30~\mu$m IR band, or $0.5$~dex for the conventional $8$--$1000~\mu$m band, and it reaches $\sim2$~dex for the least dusty hosts, which have the least dust emission for the swarm to hide behind. An IR excess that is too large for a galaxy's Balmer decrement is therefore an anomaly that ordinary dust cannot produce, and, since optical spectroscopy and WISE photometry already exist for millions of galaxies, this plane supplies an inexpensive spectroscopic pre-selection for the scaled-up searches of \S~\ref{sec:scaling}.

\begin{figure*}[!tp]
\centering
    \includegraphics[width=0.95\textwidth]{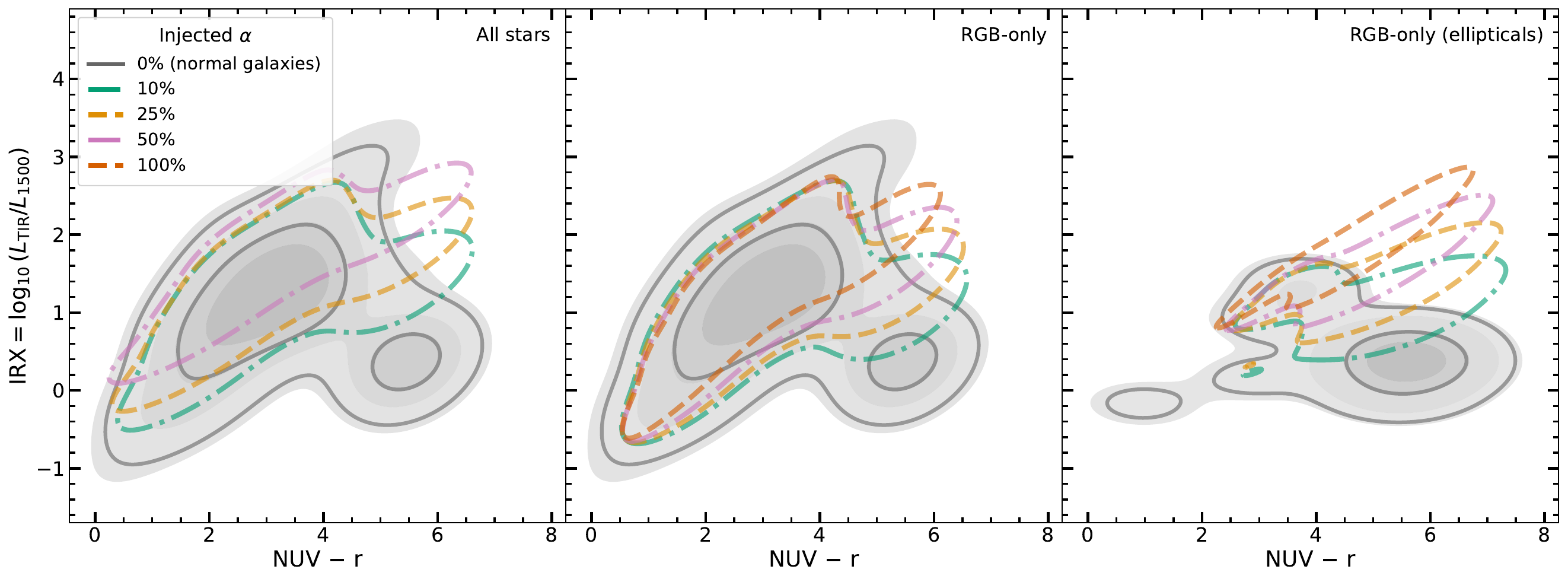}
    \caption{IR excess versus ${\rm NUV}-r$ color for the atlas under two forms of Dyson sphere injection, one acting on all stars and the other restricted to only red giant branch (RGB) stars. In each panel, the gray filled contours trace the uninjected population and the colored contours show the same galaxies with swarms injected at the labeled covering fraction. \textit{Left:} when the swarm cloaks all stars, it moves galaxies exactly vertically since the $(1-\alpha)$ dimming cancels in any color formed from bands shortward of the swarm's emission. The $100\%$ contour lies above the plotted range. \textit{Center:} swarms around red-giant-branch stars only, with $\alpha$ the covering fraction of the targeted stars. The signature is barely visible in the full atlas, whose colors are dominated by star-forming hosts. \textit{Right:} the same red-giant-only injection for the elliptical subsample, which moves upward and bluer at large covering fractions, against the direction of the observed IR-excess--color relations, since the swarms remove red light while leaving the ultraviolet untouched.}
    \label{fig:irxcolor}
\end{figure*}

A companion photometric discriminant requires no spectroscopy at all, and it extends the search to swarm architectures that do not sample the stellar population uniformly. Because a $300$K swarm emits negligibly shortward of a few microns, reprocessing all starlight uniformly acts on the UV-through-near-IR bands as a pure $(1-\alpha)$ dimming (Equation~\ref{eq:AGENT2}). This cancels exactly in any color formed from those bands, meaning that swarms acting on all stars in a galaxy fail to change the galaxy's ${\rm NUV}-r$ color. In the IR-excess versus ${\rm NUV}-r$ plane of \citet{Johnson2007} and \citet{Arnouts2013}, where ordinary attenuation moves galaxies along a sequence in which reddening and IR excess rise together, a swarm cloaking all stars therefore displaces its host exactly vertically (Figure~\ref{fig:irxcolor}, left).

A civilization that instead cloaks individual stars, as envisioned by \citet{lacki2019}, breaks this achromaticity. Using the phase-restricted variant of the formalism (\S~\ref{sec:AGENT}) to cloak the red-giant-branch stars, which supply roughly half the bolometric light of an old population, moves quiescent galaxies upward and $0.4$~mag bluer at $\alpha=1$, against the empirical dust sequence and into a region that no ordinary attenuation process populates (Figure~\ref{fig:irxcolor}, right). In the atlas as a whole, the same injection is barely visible (Figure~\ref{fig:irxcolor}, center) as young stars dominate the light of star-forming hosts, so this diagnostic, like the others in this section, is sharpest for quiescent galaxies. The displacement approaches the intrinsic color scatter of the red sequence only at large covering fractions, so we regard the plane as a population-level screen rather than a per-object test, and we note that a swarm confined to the asymptotic giant branch, which carries only a few percent of the light, would evade it entirely, leaving the near- to MIR excess as its only photometric trace. A strict luminosity threshold, cloaking every star fainter than a chosen luminosity, generalizes the same formalism to a different selection rule, and it is a natural alternative signal model for the scaled-up searches that we discuss in \S~\ref{sec:scaling}.

A similarly inexpensive diagnostic is spatial morphology, which is available from existing multiwavelength imaging. A genuine galaxy-spanning civilization would produce waste heat that traces the stellar light of the disk, whereas an AGN concentrates toward the nucleus and star-forming dust follows the spiral arms and HII regions. This is the same discriminant exploited by our resolved M77 analysis (\S~\ref{sec:resolved}). Similarly, archival diagnostic is MIR variability. The reactivated Near-Earth Object WISE mission \citep[NEOWISE;][]{Mainzer2014} surveyed the full sky in the W1 and W2 bands every six months for more than a decade, and a swarm in aggregate must be steady on these timescales, so year-to-year variability of the MIR excess points to accretion power instead. Moreover, the FIR continuum helps to distinguish swarms cooler than $\lesssim100$K from cold interstellar dust. PRIMA is well suited to this measurement, and, as a planned wide-field FIR survey, it would supply the relevant photometry for large numbers of candidates at relatively modest observational cost. Expensive diagnostics, reserved for the most compelling candidates that survive the cheaper tests, are high-resolution MIR spectroscopy and cold-dust imaging. A swarm radiates as a smooth, featureless blackbody, whereas a dusty torus exhibits silicate absorption or emission near $9.7$ and $18~\mu$m and star-forming dust exhibits PAH bands, so JWST/MIRI spectroscopy separates the two directly, while ALMA anchors the cold Rayleigh-Jeans tail. 
This ordering that we have presented here is deliberate, since the inexpensive diagnostics (i.e., archival optical spectroscopy, imaging, IR monitoring, and, eventually, wide-field PRIMA photometry), can be applied to an entire candidate list, while the costly JWST/MIRI and ALMA observations are reserved for the few candidates that remain afterwards. Since most integral field units span only a few arcseconds, nearby and spatially extended candidates may be better served by imaging combined with targeted spectroscopic pointings than by full spectral mapping, although the optimal strategy will ultimately depend on the angular size of each source.

\begin{deluxetable*}{lcccllccclcc}
\tabletypesize{\scriptsize}
\tablecaption{The ten galaxies whose real, uninjected photometry most strongly prefers a Dyson sphere component, ranked by $\Delta{\rm BIC}$ at injected $\alpha = 0$. Columns are the galaxy name, the J2000 right ascension and declination, the redshift, the Hubble type and nuclear activity class from \cite{Brown2014}, the distance, the AGENT-on $95\%$ upper limit on the covering fraction $\alpha_{95}$, $\Delta{\rm BIC} = {\rm BIC}_{\rm AGENT\,off} - {\rm BIC}_{\rm AGENT\,on}$, and checkmarks indicating existing archival coverage by JWST, ALMA, and Herschel from a $40\arcsec$ positional search of the mission archives (queried 2026 July). No galaxy reaches the $\Delta{\rm BIC}>8$ detection threshold. Table~\ref{tab:candidates} is published in its entirety in the machine-readable format, including coordinates, redshifts, recovered posterior-median covering fractions and temperatures, the JWST instrument list, and per-facility observation counts for all 129 galaxies in the sample. The ten most strongly preferred galaxies are shown here for guidance regarding its form and content.}
\label{tab:candidates}
\tablehead{\colhead{Galaxy} & \colhead{R.A. (J2000)} & \colhead{Dec. (J2000)} & \colhead{$z$} & \colhead{Type} & \colhead{Activity} & \colhead{$d$ [Mpc]} & \colhead{$\alpha_{95}$ [\%]} & \colhead{$\Delta{\rm BIC}$} & \colhead{JWST} & \colhead{ALMA} & \colhead{Herschel}}
\startdata
UGC 08696 & 13 44 42.1 & +55 53 13 & 0.0378 & Pec & AGN & 165.9 & $<26.3$ & -2.19 & $\checkmark$ & \nodata & $\checkmark$ \\
NGC 6052 & 16 05 12.8 & +20 32 32 & 0.0157 & Pec & SF & 74.8 & $<6.2$ & -5.25 & \nodata & $\checkmark$ & $\checkmark$ \\
Mrk 1490 & 14 19 43.3 & +49 14 12 & 0.0256 & Sa & SF/AGN & 115.5 & $<17.1$ & -5.51 & \nodata & \nodata & $\checkmark$ \\
UGC 08335 NW & 13 15 30.7 & +62 07 45 & 0.0308 & Pec & SF/AGN & 136.1 & $<7.5$ & -5.56 & \nodata & \nodata & $\checkmark$ \\
NGC 4594 & 12 39 59.4 & $-$11 37 23 & 0.0034 & SAa & \nodata & 9.3 & $<0.5$ & -5.70 & $\checkmark$ & $\checkmark$ & $\checkmark$ \\
NGC 6240 & 16 52 58.9 & +02 24 05 & 0.0245 & Pec & AGN & 111.9 & $<13.0$ & -5.88 & $\checkmark$ & $\checkmark$ & $\checkmark$ \\
UGC 09618 N & 14 57 00.7 & +24 37 03 & 0.0337 & Sb & SF & 151.0 & $<11.0$ & -5.94 & $\checkmark$ & $\checkmark$ & $\checkmark$ \\
NGC 7771 & 23 51 24.8 & +20 06 42 & 0.0143 & SB(s)a & SF & 60.5 & $<6.6$ & -6.10 & \nodata & $\checkmark$ & $\checkmark$ \\
IC 0883 & 13 20 35.3 & +34 08 22 & 0.0233 & Pec & SF/AGN & 106.9 & $<14.8$ & -6.11 & \nodata & $\checkmark$ & $\checkmark$ \\
NGC 4450 & 12 28 29.6 & +17 05 06 & 0.0065 & SAab & AGN & 16.5 & $<0.8$ & -6.13 & \nodata & $\checkmark$ & $\checkmark$ \\
\enddata
\end{deluxetable*}

\subsection{A Ranked Table of Anomalous MIR Emission}
\label{sec:anomalousTable}

As a first step toward a catalog of anomalous galaxies, we rank the galaxies in our atlas by the degree to which their real, uninjected photometry prefers a Dyson sphere component, adopting $\Delta{\rm BIC}$ at injected $\alpha = 0$ as the ranking statistic. This table effectively orders the galaxies in the atlas by how anomalously MIR bright they are relative to the AGENT-off model, a ranking that is useful even to studies with no interest in technosignatures. Table~\ref{tab:candidates} lists the ten most strongly preferred galaxies, and the complete ranking is available in machine-readable form. No galaxy in the atlas reaches our $\Delta{\rm BIC}>8$ detection threshold on its real photometry. In fact, no galaxy attains $\Delta{\rm BIC} > 0$. The most strongly preferred object, the merging ultraluminous IR galaxy (ULIRG) Mrk~273 (UGC~08696), attains only $\Delta{\rm BIC} = -2.2$, and the remainder of the sample lies within ${\sim}4$ units of the $-2\ln n \approx -6.4$ floor of the nested model pair, so the atlas contains no significant Dyson sphere detections.

To test how FIR data sharpen these verdicts, we measure aperture-matched \textit{Herschel} PACS photometry for the interacting merger NGC~7592 and for the strongly AGN-dominated M77 (NGC~1068), the two systems for which broadband SED fitting is least constraining. We integrate over the same \cite{Brown2014} apertures and validate against the independent photometry of the Great Observatories All-Sky LIRG Survey \citep{Chu2017}, then refit both galaxies with and without a Dyson sphere component. For NGC~7592, the PACS points anchor the cold-dust continuum and confine the AGENT-on component to $T_{\rm BB} = 53_{-5}^{+9}$K, the temperature of ordinary interstellar dust rather than of plausible waste heat, while the model selection remains null ($\Delta{\rm BIC} = -6.3$ to $-6.4$). For M77, whose degeneracy lies in the mid IR beyond the reach of PACS, both the model selection ($\Delta{\rm BIC} = -6.4$ in both cases) and the covering-fraction limit ($\alpha_{95} = 0.31$ to $0.32$) are insensitive to the added bands, as expected. Together, the two cases show that follow-up photometry vets a candidate by measuring the temperature of its residual excess, and that the constraint must be sought where the degeneracy lives, in the far IR for cold-dust impostors and in the mid IR for AGN-dominated systems.

This null result is expected considering the fact that these are all nearby and well-studied targets, so it is reassuring that the AGENT-on covering fractions that we recover for even the most preferred objects remain modest. We note that the galaxies with the largest BIC include the same IR-luminous interacting mergers and dust-obscured AGN that drive our injection-recovery outliers, Mrk~273, NGC~6240, and Mrk~1490 among them, rather than quiescent galaxies with otherwise unexplained MIR excesses. These IR-bright mergers are nonetheless the most appropriate first targets for the confirmatory spectroscopy described above, since they are the objects for which integrated-light SED fitting is least able to exclude a buried waste-heat component. In this sense Table~\ref{tab:candidates} defines a target list of galaxies whose broadband photometry alone cannot rule out a Dyson sphere, and it is this list, rather than any claim of a detection, that motivates the dedicated follow-up campaigns outlined above.

\section{Conclusions}
\label{sec:conclusions}

Here, we present the most robust stellar population synthesis-based search for galaxy-spanning technological waste heat conducted to date, incorporating the AGENT formalism of \cite{Wright2014a} into \texttt{FSPS} at the stellar population level and pairing the resulting forward model with the \prospector Bayesian inference framework. We fit 1,419 injected SEDs and the real photometry of all 129 galaxies in the \cite{Brown2014} atlas, each with and without AGENT parameters. Our principal conclusions are as follows.

\begin{itemize}
    \item Waste heat injected at the SSP level propagates self-consistently into the nebular and dust emission of the model galaxy, and our injection-recovery campaign validates the full pipeline. That is, across all 129 galaxies and 11 injection levels we recover the injected covering fractions with a best-fit slope of $m = 0.92$, and no galaxy is spuriously detected at zero injection.

    \item Detectability is easiest in the outskirts of quiescent galaxies. Bayesian model selection recovers covering fractions as low as ${\sim}4$--$5\%$ in old, quiescent galaxies, whereas dusty star-forming systems require $\alpha \gtrsim 20\%$. Restricting the fits to wide-field survey photometry raises these thresholds by a further factor of ${\sim}3$--$4$ for quiescent galaxies, so the UV-through-MIR coverage assembled here is essential to the limits we report.

    \item None of the 129 galaxies prefer a Dyson sphere component on its real, uninjected photometry. The AGENT-on fits instead place the first per-galaxy 95\% upper limits on the covering fraction, with a median $\alpha_{95}$ of $0.3\%$ across quiescent hosts without a dominant AGN and $89\%$ of them below $3\%$, and the ranking of the full sample by $\Delta{\rm BIC}$ (Table~\ref{tab:candidates}) defines the target list for follow-up. Combining the zero detections with the injection-calibrated detection counts also bounds the population, so fewer than $14.6\%$ of atlas-like galaxies can host $300$K swarms intercepting $5\%$ of their starlight, and fewer than $2.6\%$ can host swarms intercepting $25\%$, each at $95\%$ confidence.

    \item When waste heat is present in a galaxy's photometry but absent from the model, the fitter compensates in two falsifiable ways. That is, it inflates the inferred AGN fraction by up to three orders of magnitude and the recent star formation rates of quiescent galaxies by $1.4$--$1.8$~dex, so an anomalously luminous AGN posterior or an unphysically elevated recent SFH in an otherwise unremarkable galaxy is a practical red flag for unmodeled MIR emission.

    \item The dominant systematic is the degeneracy between warm AGN dust and waste heat, and spatially resolved fitting with nuclear excision breaks it. For M77, excising the inner $34\arcsec$ lowers the recovered covering fraction from $\alpha = 0.182$ to $\alpha = 0.100_{-0.054}^{+0.137}$, consistent with zero at $1.9\sigma$ and with the $\sim5\%$ of AGN light that leaks past the excision radius.

    \item Future observations extend these limits in complementary ways. Our mock-photometry experiments show that PRIMAger photometry pins the FIR continua of the waste-heat and AGN hypotheses jointly, concentrating their remaining disagreement in the $8$--$13~\mu$m region that JWST/MIRI can test directly. Archival optical spectroscopy vets candidates at essentially no cost through AGN emission lines, the Balmer decrement, and dynamical masses. The invariance of ultraviolet-optical colors under reprocessing of all starlight turns the IR-excess versus color plane into a further archival screen, one that extends to swarms cloaking individual stars. Simulation-based inference (L. Shi et al., in preparation) reduces the per-galaxy cost of these fits by five orders of magnitude, opening the $\sim10^7$ galaxies with archival multiwavelength photometry to this search.
\end{itemize}

Here, we necessarily focus on a small, well-characterized atlas of nearby galaxies, and the limits we report are individually modest. The machinery this paper establishes is nevertheless precisely what a survey-scale search requires, namely a validated forward model for waste heat, calibrated per-galaxy posteriors, characteristic failure modes of waste-heat-free models, and a tiered confirmation strategy. Old, quiescent galaxies, where every natural source of MIR emission is faint and predictable, are the best places to look for technological waste heat, and their outskirts, beyond the reach of AGN contamination, are the natural place to focus on.

\section{Acknowledgments}
O.C. acknowledges support as a Penn State Extraterrestrial Intelligence Center Postdoctoral Fellow, which is funded through a private donation made by The Ultraintelligence Foundation. 

This research has made use of NASA's Astrophysics Data System Bibliographic Services. This research was partially supported by the Seed Grant award ICDS\_READ26\_m1jtw13 from Penn State’s Institute for Computational and Data Sciences. The Penn State Extraterrestrial Intelligence Center and the Center for Exoplanets and Habitable Worlds are supported by Penn State and its Eberly College of Science. 

J.M.H. acknowledges support from the Evolving Universe Fellowship, which is made possible by a generous donation from Dr. Keiko Miwa Ross. J.M.H. also acknowledges support from JWST Program \#8544.

\section{Data and Software Availability}
\label{sec:dataavail}

The photometry fitted in this work is the published matched-aperture atlas of \cite{Brown2014}, whose data release is hosted as a High Level Science Product on the Mikulski Archive for Space Telescopes (MAST): \dataset[10.17909/t9-5bxk-dh29]{https://doi.org/10.17909/t9-5bxk-dh29}. The \textit{Herschel} PACS measurements of \S~\ref{sec:confirming} were reduced from public archival imaging. The full ranking of all 129 galaxies, with coordinates, redshifts, distances, morphological and activity classes, $\Delta{\rm BIC}$, posterior-median and $95\%$ upper-limit covering fractions, recovered blackbody temperatures, AGN fractions, and per-facility archival observation counts, is published in machine-readable form as Table~\ref{tab:candidates}.

The AGENT implementation is distributed as public forks of the three codes it modifies, namely \texttt{FSPS}\footnote{\url{https://github.com/o-curtis/fsps}}, \texttt{python-fsps}\footnote{\url{https://github.com/o-curtis/python-fsps}}, and \prospector\footnote{\url{https://github.com/o-curtis/prospector}}. The \texttt{FSPS} fork adds the AGENT reprocessing of Equation~\ref{eq:AGENT2} at the simple stellar population level together with the phase-restricted variant of \S~\ref{sec:AGENT}, the \texttt{python-fsps} fork exposes \DSalpha and \DSTBB through the Python interface, and the \prospector fork supplies the model templates and parameter transforms that make them free parameters of the fit. Installing all three reproduces the forward model used throughout this work.

\facilities{GALEX, Swift(UVOT), Sloan, FLWO:2MASS, CTIO:2MASS, WISE, NEOWISE, Spitzer, Herschel(PACS)}

\software{\texttt{FSPS} \citep{Conroy2009},
          \prospector \citep{leja2017, Johnson2021},
          \texttt{emcee} \citep{ForemanMackey2013},
          \texttt{dynesty} \citep{Speagle2020},
          \texttt{piXedfit} \citep{abdurrouf2021},
          \texttt{Astropy} \citep{astropy2013, astropy2018, astropy2022},
          \texttt{NumPy} \citep{numpy2020},
          \texttt{SciPy} \citep{scipy2020},
          \texttt{Matplotlib} \citep{matplotlib2007}}

\appendix
\restartappendixnumbering

\section{Quantifying the Temperature Dependence of the Waste-Heat Degeneracy}
\label{sec:Tappendix}

Our choice of the fiducial $T_{\rm BB} = 300$K in \S~\ref{sec:galCat} follows from where the collectors that gather starlight are expected to sit. A swarm element that intercepts light at roughly habitable-zone distances from a Sun-like star radiates at an equilibrium temperature of a few hundred kelvin, which is close to Dyson's own original estimate \citep{Dyson1960}, and a $300$K blackbody peaks near $\sim10~\mu$m, squarely within the WISE W3 and WISE W4 bands where our reddest photometry lies. A separate and often-conflated argument sets how cold a civilization could in principle radiate its waste heat. A society optimizing for thermodynamic efficiency could dispose of energy at temperatures approaching the cosmic microwave background \citep{Wright2023}, far colder than $300$K, so our fiducial value should be read as a conservative and observationally convenient midpoint rather than a hard prediction.

The swarm temperature also sets which astrophysical component is most easily confused with the waste heat, as the bottom panel of Figure~\ref{fig:FSPSTest} makes clear. At hot temperatures of $500$--$1000$K the blackbody peaks at $3$--$6~\mu$m, where hot AGN-torus dust and the Rayleigh-Jeans tail of the stellar photosphere both contribute, so both the AGN degeneracy and a new degeneracy with starlight worsen. At cold temperatures of $50$--$150$K, the peak moves into the far IR beyond $30~\mu$m where the AGN torus contributes little but cold interstellar dust dominates, so the principal confounder shifts from the AGN to ordinary dust. Our fiducial $300$K sits in a relatively favorable window near $10~\mu$m between these two regimes, still overlapping warm AGN dust but not yet buried in either the stellar tail or the cold-dust peak. This argument is qualitative and made from the shape of the injected spectra alone, so the remainder of this appendix describes a dedicated injection-recovery experiment designed to turn it into a measurement.

We select 15 galaxies from the \cite{Brown2014} atlas that span the morphological range, including quiescent ellipticals, star-forming spirals, and IR-luminous interacting mergers. For each galaxy, we inject waste heat at three blackbody temperatures, $T_{\rm BB} = 100$, $300$, and $600$K, across the same covering-fraction grid used in the main analysis, $\alpha \in \{0, 1, 2, 3, 4, 5, 10, 15, 20, 25, 50\}\%$. The injected photometry is generated with the same AGENT-enabled forward model used throughout this work, so that each injected SED is a self-consistent reprocessing of the galaxy's own best-fit spectrum rather than an analytic addition to the broadband points. We then fit every injected SED both with and without the AGENT parameters using the \texttt{emcee} sampler with 256 walkers and 15,000 steps, exactly as in \S~\ref{sec:bayes}, and we calculate likelihoods for our BIC tests exactly as we describe in the main text. The $300$K fits are those already obtained in \S~\ref{sec:injectRecov}, while the $100$ and $600$K fits are run for this dedicated subsample.

\begin{figure*}[!tp]
    \centering
    \includegraphics[width=0.72\textwidth]{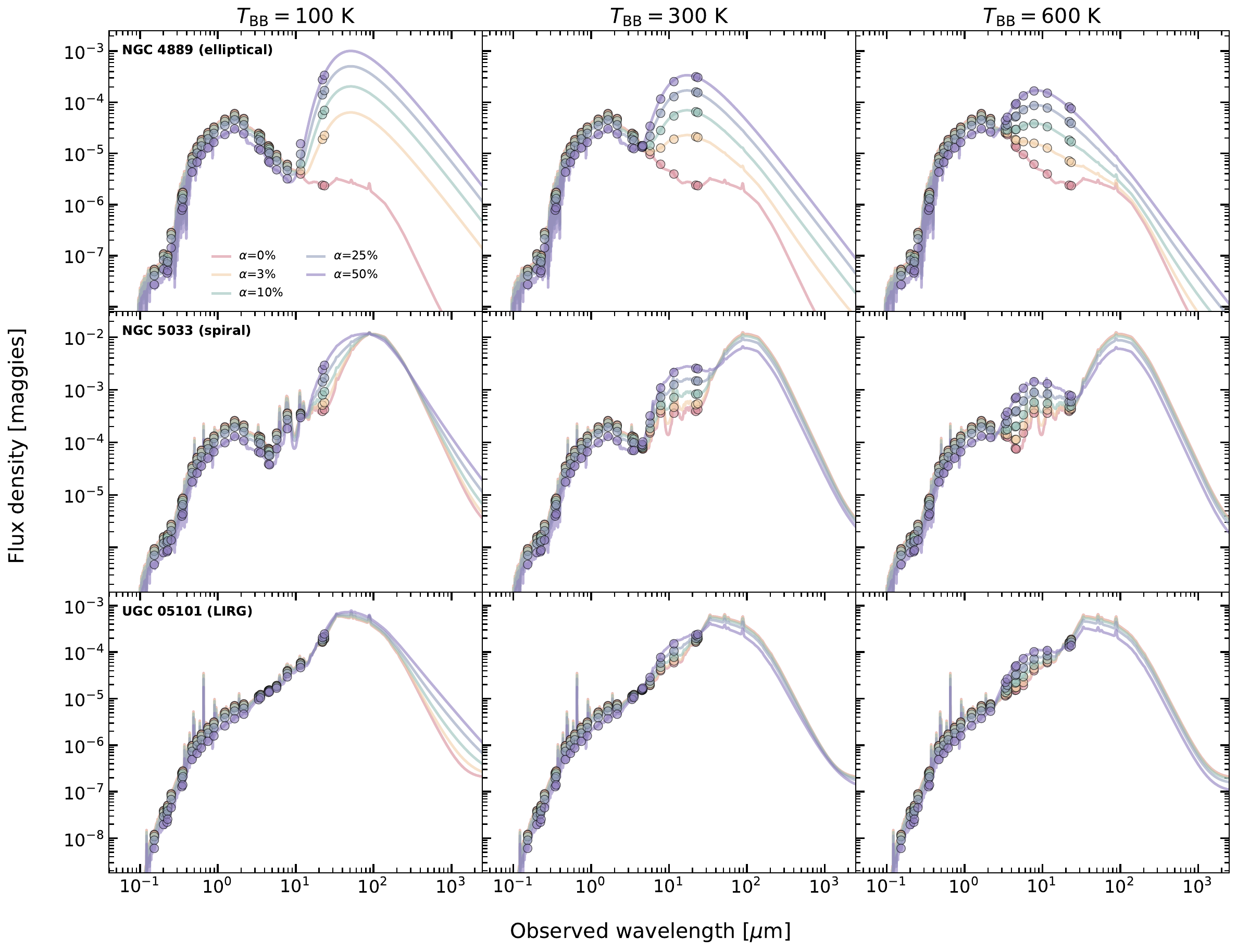}
    \caption{Injected spectral energy distributions for three representative galaxies, a quiescent elliptical (top), a star-forming spiral (middle), and an IR-luminous merger (bottom), at $T_{\rm BB} = 100$, $300$, and $600$K (left to right). In each panel, the solid curves show the AGENT-reprocessed model spectra at covering fractions $\alpha = 0$, $3$, $10$, $25$, and $50\%$, and the filled circles show the injected broadband photometry. The injected blackbody peak marches from roughly $40~\mu$m at $100$K, through $\sim15~\mu$m at $300$K, to $\sim6~\mu$m at $600$K, while its amplitude grows with $\alpha$ and the UV and optical continuum dims as $(1-\alpha)$. These injected SEDs make the two-sided nature of the signal explicit, in that the same covering fraction that builds the MIR bump also dims the UV and optical continuum, and it is that pairing which full-SED fitting exploits.}
    \label{fig:injectverify}
\end{figure*}

Figure~\ref{fig:injectverify} shows the injected SEDs for three representative galaxies at the three temperatures, with the injected broadband photometry overplotted on the model spectra. The behavior anticipated in \S~\ref{sec:galCat} is clearly visible. The injected blackbody peak marches from roughly $40~\mu$m at $100$K, through $\sim15~\mu$m at $300$K, to $\sim6~\mu$m at $600$K, so that a cold swarm deposits its excess in the far IR where cold interstellar dust dominates, whereas a hot swarm deposits it in the near-to-mid IR where the warm AGN torus and the stellar continuum both contribute. The figure also illustrates why the recoverability depends on morphology. In the quiescent elliptical the injected bump dominates the otherwise faint IR and stands out at every temperature, whereas in the IR-luminous merger the same fractional injection is a far smaller perturbation on the strong star-forming dust peak.

These fits quantify how the degeneracy strength varies with temperature through three diagnostics, namely the scatter in the recovered covering fraction relative to the injected value, the model-selection threshold at which a $\Delta{\rm BIC} > 8$ detection becomes likely, and the inflation of the inferred AGN fraction and torus optical depth in the AGENT-off fits. Following the argument of \S~\ref{sec:galCat}, we expect that the recovered-covering-fraction scatter and the AGN-parameter inflation to be largest at $600$K, where the waste heat is most readily absorbed by the warm AGN component, to be intermediate at the fiducial $300$K, and to migrate out of the AGN parameters and into the cold-dust parameters at $100$K, where the excess falls in the far IR.

\begin{figure*}[!tp]
    \centering
    \includegraphics[width=0.58\textwidth]{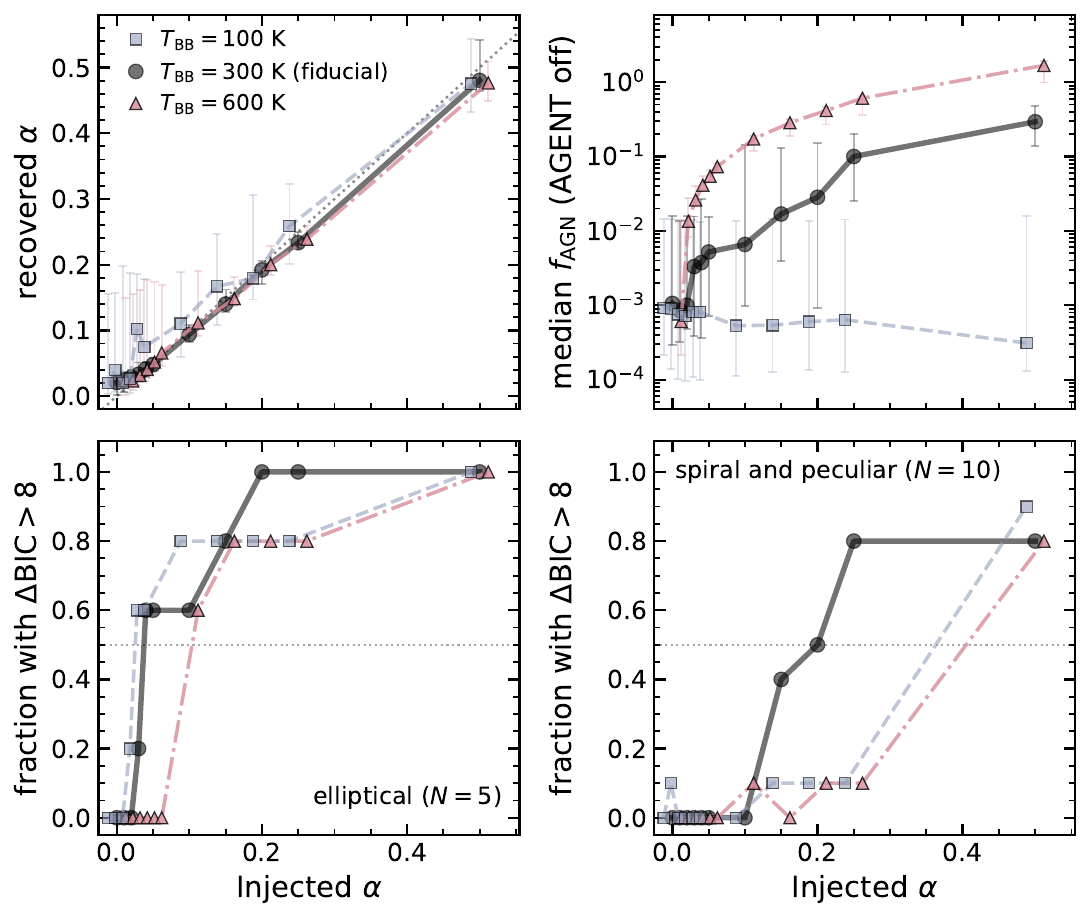}
    \caption{Results of the temperature-dependence experiment for the fifteen-galaxy subsample, with injected swarm temperatures of $100$K (pale blue squares, dashed), $300$K (black circles, solid), and $600$K (rose triangles, dash-dotted). \textbf{Top left:} median recovered covering fraction versus injected covering fraction, with error bars spanning the 16--84\% range across the subsample and the dashed line marking the 1:1 relation. Recovery is unbiased at all three temperatures. \textbf{Top right:} population-median AGN fraction inferred by the AGENT-off fits, with error bars spanning the 16--84\% range across the subsample. Hot injections inflate the inferred AGN fraction by up to three orders of magnitude, whereas cold injections leave it untouched because their excess is absorbed by the cold-dust parameters instead. \textbf{Bottom:} fraction of galaxies with $\Delta{\rm BIC} > 8$ as a function of injected covering fraction, shown separately for the five ellipticals (left) and the ten spirals and peculiars (right). We note that this split is made on the numerical $T$-type of \cite{Brown2014}, which groups the two early-type peculiars in the subsample (Arp~118 and NGC~0750), with the ellipticals, whereas Figures~\ref{fig:DSaVsTruth} and~\ref{fig:limitedFilterResults} and the machine-readable version of Table~\ref{tab:candidates} bin on the morphological type string. This distinction does not strongly affect the results of this Appendix. The ellipticals are detected efficiently at every temperature, while the dusty galaxies carry the entire temperature dependence and are detectable at low covering fractions only near the fiducial $300$K.}
    \label{fig:Tresults}
\end{figure*}

Figure~\ref{fig:Tresults} presents the outcome of the experiment. The recovery machinery itself is robust across the full temperature range, as the top left panel shows. The recovered covering fractions track the injected values with median offsets below $\sim0.03$ for $\alpha \geq 10\%$ at all three temperatures, no galaxy is spuriously detected at $\alpha = 0$ at any temperature, and the injected temperature is itself recovered without bias, with median recovered \DSTBB\ of $97$, $301$, and $600$K at $\alpha = 25\%$. What changes with temperature is therefore not the fidelity of the posteriors but the difficulty of the model-selection problem.

The bottom panels quantify that difficulty and separate it by morphology. For the ellipticals, the detection efficiency is nearly independent of temperature, with the majority of the subsample detected by $\alpha \approx 5$--$10\%$ at all three temperatures, as these galaxies have so little competing IR emission that an excess at any of the three temperatures is anomalous. The temperature dependence is instead concentrated in the star-forming spirals and peculiars. There, the fiducial $300$K case reaches $80\%$ detection by $\alpha = 25\%$, whereas the $100$ and $600$K cases remain nearly undetectable until $\alpha \approx 50\%$. The two regimes fail for different reasons. A $100$K blackbody peaks near $40~\mu$m, where cold interstellar dust provides a natural absorber and only the Wien side of the bump enters our reddest bands, whereas a $600$K blackbody peaks near $6~\mu$m, where the stellar Rayleigh-Jeans tail and the warm torus supply equally natural absorbers. A $300$K swarm instead peaks near $10$--$15~\mu$m, in the gap between these astrophysical components, where the WISE W3 and W4 and \textit{Spitzer} $24~\mu$m bands retain full leverage. The fiducial temperature therefore sits in a genuinely favorable window for the dusty galaxies that dominate the atlas, and the detection thresholds reported in \S~\ref{sec:modelSelection} should be read as optimistic relative to swarms much hotter or colder than a few hundred kelvin.

The top right panel shows where the unmodeled excess goes in the AGENT-off fits, and it confirms the expectation of \S~\ref{sec:galCat} in detail. At $600$K, the population-median $f_{\rm AGN}$ inflates by more than three orders of magnitude between $\alpha = 0$ and $\alpha = 50\%$, roughly an order of magnitude more inflation than the fiducial $300$K case. At $100$K, the AGN fraction never moves, holding at its baseline of a few $\times10^{-4}$ at every injection level since the cold excess is absorbed by the cold interstellar dust parameters instead. The red-flag diagnostics of \S~\ref{sec:AGNParams} are therefore temperature specific, and an anomalously luminous AGN posterior signals a hot swarm, whereas a cold swarm would instead masquerade as an anomalously luminous cold-dust component.

\section{Spectroscopic and Dynamical Discriminants}
\label{sec:massAppendix}
\setcounter{figure}{0}

The tiered confirmation strategy of \S~\ref{sec:confirming} draws two of its least expensive diagnostics from a single archival optical spectrum, the Balmer decrement and the dynamical mass. Whether either is decisive or merely corroborating for a given candidate depends on how cleanly the presence of a swarm separates the true interpretation of that galaxy's spectrum from the dusty, swarm-free interpretation an analyst would otherwise adopt, and our injection suite measures that separation directly since all of our injected spectra are fit both with the AGENT parameters turned on and off. This appendix uses those paired fits to calibrate the discriminating power of each test against the known truth, which is what sets the tiering adopted in \S~\ref{sec:confirming}.

\begin{figure*}[!tp]
    \centering
    \includegraphics[width=0.66\textwidth]{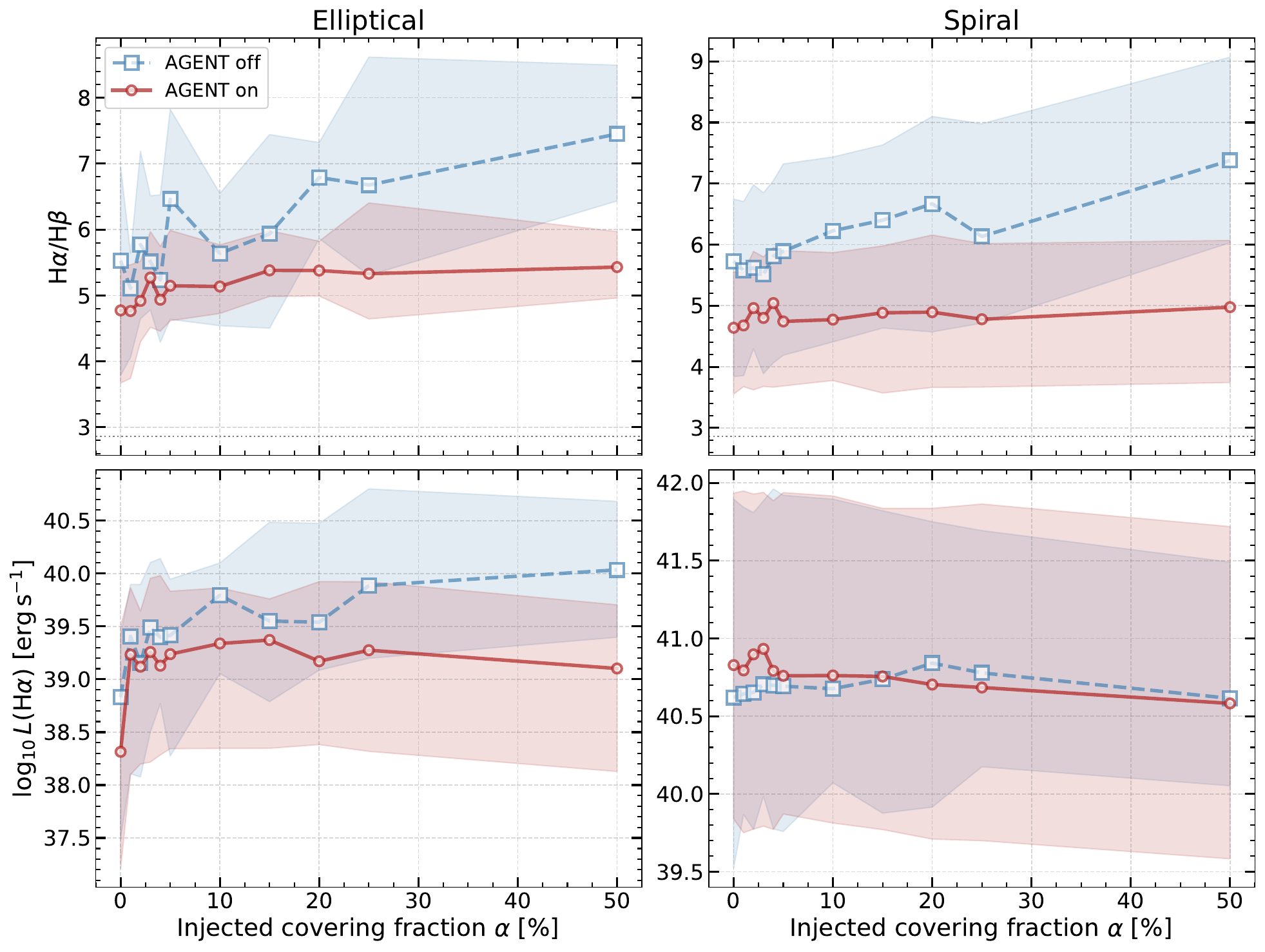}
    \caption{Posterior-predictive Balmer decrement (top) and H$\alpha$ luminosity (bottom) of the AGENT-on (red circles) and the AGENT-off fits (blue squares), against injected covering fraction for ellipticals (left) and spirals (right). Points are population medians and the shaded regions span the inner 64\% range of the data. Forced to reproduce the injected MIR excess without a swarm, the AGENT off models invoke extra dust and star formation, over-predicting the nebular reddening, and, for ellipticals, the H$\alpha$ luminosity. The decrement separates the two interpretations for both morphologies, whereas the H$\alpha$ luminosity is degenerate with ordinary star formation in spirals.}
    \label{fig:balmergrid}
\end{figure*}

The Balmer decrement follows from the same logic as the AGN and star-formation failure modes of \S~\ref{sec:AGNParams}. Forced to reproduce a cloaked galaxy's MIR excess with dust and star formation rather than a swarm, the AGENT-off model invokes more nebular reddening and extra ionizing photons, so it predicts a stronger Balmer decrement and a brighter H$\alpha$ line than the truth. We quantify this by drawing 40 posterior samples from every fit, predicting the full model spectrum, and measuring the emission-line luminosities from the difference between spectra computed with and without nebular emission, which isolates the lines from the underlying stellar Balmer absorption. Figure~\ref{fig:balmergrid} shows the resulting posterior-predictive Balmer decrement and H$\alpha$ luminosity of both fits against the injected covering fraction. At the population level, the separation is clean; when the AGENT parameters are turned off, the median increase from the true value of ${\sim}5$ to ${\sim}7.5$ when $\alpha = 50\%$ for both morphologies, and it inflates the luminosity of H$\alpha$ of the ellipticals while the H$\alpha$ excess is masked by the genuine ongoing star formation within spiral galaxies. A galaxy whose measured decrement is far below the value that the dusty interpretation requires is therefore incompatible with that interpretation.

For a single object, the discrimination is more modest. After propagating each fit's posterior-predictive spread with the ${\sim}5\%$ per-line errors of a typical archival spectrum, the significance with which one decrement measurement separates the two interpretations rises with covering fraction but stays below ${\sim}1.5\sigma$ over the range we probe, reaching a median $N_\sigma \simeq 0.9$ for ellipticals and $1.1$ for spirals from the decrement, and $1.5$ for ellipticals from the H$\alpha$ luminosity, at injected $\alpha = 50\%$. The dusty fit predicts an inflated decrement, but with a broad enough posterior that a single spectrum rarely excludes it outright. The Balmer decrement is thus a decisive discriminant in aggregate and a corroborating one for an individual candidate, the same tiering we find below for dynamical masses.

\begin{figure*}[!tp]
    \centering
    \includegraphics[width=0.78\textwidth]{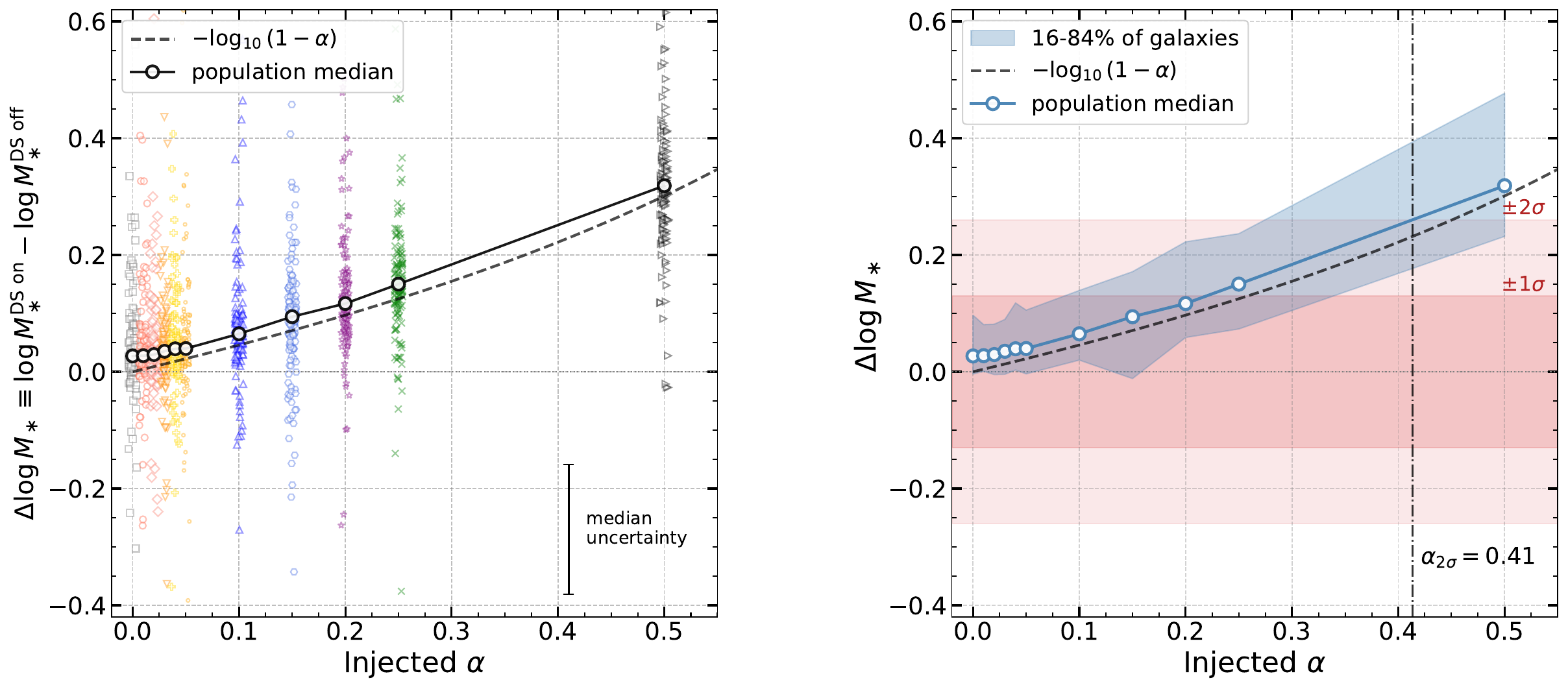}
    \caption{The hidden-mass offset between the two interpretations of every injected SED. \textbf{Left:} difference between the posterior median stellar masses of the AGENT-on and AGENT-off fits for all 1,419 injection pairs, colored by injection level as in Figure~\ref{fig:DSaVsTruth}, with the population median (open circles) tracking the $-\log_{10}(1-\alpha)$ prediction (dashed). \textbf{Right:} the same median with its 16--84\% population range (shaded) against the $\pm1\sigma$ and $\pm2\sigma$ observed scatter of the stellar-mass--dynamical-mass relation \citep[$0.13$~dex,][]{Taylor2010}. The offset becomes a $2\sigma$ dynamical-mass discriminant only for injected $\alpha \geq 0.41$ (vertical line).}
    \label{fig:hiddenmass}
\end{figure*}

A Dyson swarm also hides starlight without removing stellar mass, so the two interpretations of a cloaked galaxy disagree about how much stellar mass is present. The AGENT-off interpretation must explain the dimmed UV--optical SED with less mass, underestimating $\log M_\ast$ by $-\log_{10}(1-\alpha)$, whereas the AGENT-on interpretation recovers the true value. Our injection suite measures this offset directly since every injected SED is fit with both models (\S~\ref{sec:injectRecov}). The left panel of Figure~\ref{fig:hiddenmass} shows the difference between the posterior median stellar masses of the two fits for all 1,419 AGENT-on/off pairs. The population median tracks the $-\log_{10}(1-\alpha)$ prediction at every injection level, reaching $+0.32$~dex at injected $\alpha = 50\%$ against a predicted $+0.30$~dex, and the offset is carried almost entirely by the AGENT-off masses, which fall below their uninjected baselines by $-0.33$~dex at injected $\alpha = 50\%$ while the AGENT-on masses drift by no more than $0.03$~dex.

The right panel compares this hidden-mass signal with the precision of the dynamical masses that would test it. The observed scatter between stellar masses and structure-corrected dynamical masses in the local universe is $0.13$~dex \citep{Taylor2010}, before any allowance for the systematic variation of the stellar initial mass function seen in early-type galaxies \citep{Cappellari2013}, and the population-median offset exceeds twice that scatter only at the crossing marked at $\alpha_{2\sigma} = 0.41$, which moves to $\alpha \approx 0.45$ once the stellar-mass posterior width is folded into the error budget. A dynamical mass is therefore a decisive discriminant only for the most heavily cloaked galaxies, and at the few-percent covering fractions to which our search is sensitive it can provide strong supporting evidence. The same offset doubles as a failure-mode diagnostic in the sense of \S~\ref{sec:AGNParams} since the AGENT-off interpretation of a genuinely cloaked galaxy is wrong about its stellar mass in addition to its dust and AGN content.

\bibliography{sample701}{}
\bibliographystyle{aasjournalv7}

\end{document}